\documentclass[useAMS,usenatbib]{mn2e}
\usepackage{graphicx}
\usepackage{rotating}
\usepackage{lscape}

\title[The STREGA survey]{The STREGA survey. II. Globular Cluster
  Palomar 12\thanks{In memory of our beloved colleague and friend
    Geppina Coppola. Based on data
    collected with the ESO INAF - VLT Survey Telescope and OmegaCAM at
    the European Southern Observatory, Chile (ESO Programme
    091.D-0623, 093.D-0170)}}
\author[Musella et al.]{I. Musella$^{1}$\thanks{E-mail: ilaria.musella@oacn.inaf.it},
 M. Di Criscienzo$^{2}$,  M. Marconi$^{1}$, G. Raimondo$^3$,  V. Ripepi$^1$,
\newauthor
M. Cignoni$^{4}$, G. Bono$^{5}$,  E. Brocato$^2$, M. Dall'Ora$^{1}$, I. Ferraro$^{2}$, A. Grado$^1$,
\newauthor
G. Iannicola$^{2}$,
  L. Limatola$^1$,  R. Molinaro$^1$, M. I. Moretti$^1$, P. B. Stetson$^6$,
\newauthor
  M. Capaccioli$^{7}$, M.-R.L. Cioni$^{8,9}$,
 F. Getman$^{1}$,
and P. Schipani$^1$\\
$^{1}$INAF-Osservatorio Astronomico di Capodimonte, Salita
  Moiariello, 16, I-80131, Napoli, Italy\\
$^{2}$INAF-Osservatorio Astronomico di Roma, Via Frascati 33, I-00044
Monte Porzio Catone, Italy\\
$^{3}$INAF-Osservatorio Astronomico Collurania, via M. Maggini, I-64100 Teramo, Italy\\
$^{4}$Dipartimento di Fisica ``Enrico Fermi'', Universit\`a di Pisa, largo Pontecorvo 3, 56127, Pisa, Italy\\
$^{5}$Dipartimento di Fisica, Universit\`a degli Studi di Roma-Tor Vergata, via della Ricerca Scientifica 1, I-00133 Roma, Italy \\
$^{6}$ NRC Herzberg Astronomy and Astrophysics, 5071 West Saanich Road, Victoria, BC V9E 2E7, Canada\\
$^{7}$Dipartimento di Fisica, Universit\`a ``Federico II'', Via
Cinthia, I-80126 Napoli, Italy\\
$^{8}$Leibnitz-Institut f\"ur Astrophysik Potsdam, An der Sternwarte 16, D-14482 Potsdam, Germany\\
$^{9}$University of Hertfordshire, Physics Astronomy and Mathematics, 
College Lane, Hatfield AL10 9AB, UK\\
}
\begin{document}

\date{Accepted ..... Received .... ;}


\maketitle

\label{firstpage}

\begin{abstract} In the framework of the STREGA (STRucture and
Evolution of the GAlaxy) survey, two fields around the globular
cluster Pal~12 were observed with the aim of detecting the possible
presence of streams and/or an extended halo. The adopted stellar
tracers are the Main Sequence, Turn-off and Red Giant Branch stars. We
discuss the luminosity function and the star counts in the observed
region covering about 2 tidal radii, confirming that Pal~12 appears to
be embedded in the Sagittarius Stream. Adopting an original approach
to separate cluster and field stars, we do not find any evidence of
significant extra-tidal Pal~12 stellar populations. The presence of
the Sagittarius stream seems to have mimicked a larger tidal radius in
previous studies. Indeed, adopting a King model, a redetermination of
this value gives $r_T=0.22\pm0.1$ deg.
\end{abstract}


\section{Introduction}

This paper is the second of a series devoted to the results of the
STREGA (STRucture and Evolution of the GAlaxy) survey
\citep{Marconi14}.  This is a guaranteed-time project that uses the
ESO Very Large Telescope Survey Telescope (VST) to observe extensive
regions around a number of globular clusters (GCs) and satellite
galaxies of the Milky Way, in order to map the existence of extended
haloes and/or tidal streams with the final aim of constraining the
formation and evolution of the Galactic halo.  A detailed
description of this survey is reported in \citet{Marconi14}.  In this
paper, we describe the results we obtain by exploring a region of more
than 1 square degree around the GC Pal~12. This target is located at
(l,b)=(30,51, 47.68 deg), it has an absolute visual magnitude $M_V=
-4.47$ mag, a half-light radius of 1.72 arcmin ($=0.03$ deg), a tidal
radius $r_T$ and a core radius $r_C$, obtained on the basis of a
King model with a central concentration $c=log(r_T/r_C)$
\citep{Harris96,McL05}, of 17.42 arcmin ($=$ 0.29 deg) and 0.02 arcmin
($=0.0003$ deg, with $c=2.98$), respectively. It is worth noting that
\citet{VDB11} pointed out that the small stellar sample
contained in the cluster and the presence of substructures might cause
errors in the measured central concentration and in the determination
of the structural radii.  This GC is probably younger and more metal
rich than the majority of the Galactic GCs (GGCs)
\citep{GO88,Stetson89,Rosenberg98}. On this basis, many authors have
suggested the possibility that this GC was accreted from a
surrounding galaxy such as, for example, the Magellanic Clouds
\citep[e.g.][and references therein]{Lin92,Zinn93}. Conversely,
\citet{Irwin99} pointed out that distance and radial velocity of
Pal~12 are consistent with the hypothesis that it has been captured by
our Galaxy in a tidal interaction with the Sagittarius dwarf
Spheroidal (Sgr dSph) galaxy. The latter hypothesis was supported by
\citet{Dinescu00} through the determination of the proper motions and
the three-dimensional orbit of Pal~12.   The presence of an additional stellar
population in the direction of this GC was detected by \citet{MD02}
analysing a large field around Pal~12 and by
\citet{Bellazzini03} using data from the 
2-Micron All-Sky Survey. This population appears to be at the same
distance (within the uncertainties), but more
metal poor than Pal~12, even if with a significant spread in
metallicity and/or age, as expected for a dSph galaxy. This stellar
population was confirmed to be a part of the Sgr stream through a
comparison of its CMD with that of the Sgr dSph main body and/or with
previous CMDs of portions of the Sgr southern stream \citep[see
e.g.][and references therein]{Majewski99,Newberg02}.

In this paper, exploiting the potentiality of the large field of view
of VST, we study the stellar populations around this cluster to
better investigate its properties and the extent of its halo,
taking into account the already known presence of the Sgr stream.

In section \ref{sec-red}, the observations and the data reduction
procedures are described. An analysis of the Pal~12 CMD is presented
in Section \ref{sec-cmd}  and the luminosity function and the star
counts around this cluster are discussed in Sections \ref{sec-lum} and
\ref{sec-counts}, respectively.  The summary closes the paper.

\section{Observations and data reduction}\label{sec-red}

In the first paper of the STREGA survey \citep{Marconi14}, we
presented preliminary results obtained for one field centered on
the GC Pal~12 (F1). In this field, a large part of the cluster fell in the
central gap of the VST camera OMEGACAM, and for this reason we
subsequently observed another field (F2) with Pal~12 centered in one
of the CCDs.

\begin{figure}
\includegraphics[width=0.45\textwidth]{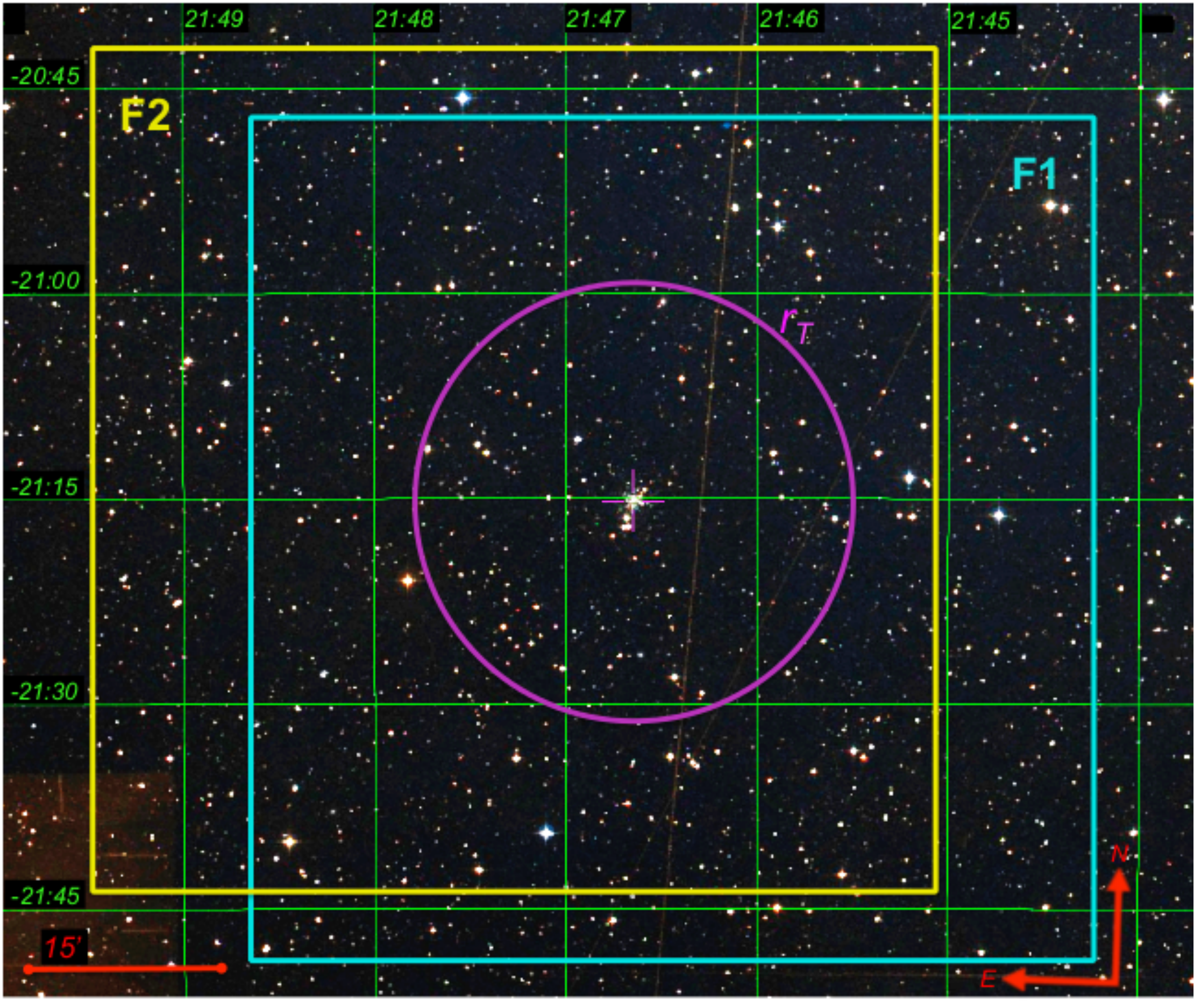}
\caption{A DSS Image of the region around Pal~12 covered by the two selected VST
  fields labelled as F1 and F2. The magenta circle shows the area
  included in the tidal radius of 0.29 deg.}
\label{region}
\end{figure}

The final observed region is shown in Fig. \ref{region}. The
observations of the F1 and F2 fields were performed on 2013-07-1/2 and
2014-07-21/22, with exposure times of 170$\,$s, 160$\,$s and 405$\,$s for the $g$,
$r$ and $i$ bands, respectively.

The data reduction was carried out with the VSTTube imaging pipeline
\citep{grado12}. The detailed procedure is described in \citet{Marconi14}.

To derive stellar photometry, we used the DAOPHOT/ALLSTAR
\citep{Stetson87} programs that allowed us to perform point-spread
function (PSF) fitting method to obtain accurate
photometry even for the stars in the crowded center of Pal~12. First,
we combined the photometry of the two fields in each band, assuming
for the stars in the overlapping region with two different magnitude
estimates, an average of the two values weighted with the photometric
uncertainties. To minimize the nonstellar or spurious
objects, the final $gri$ catalogue include only the stars
with $\chi<1.5$\footnote{The DAOPHOT/ALLSTAR $\chi$ parameter of each
star is a robust estimate of the ratio of the observed pixel-to-pixel scatter of
the fitting residuals to the expected scatter. The $\chi$ in the final
catalogue is an average of the $\chi$ values in the single frame and
single band file.}.

\begin{figure}
\includegraphics[width=0.45\textwidth]{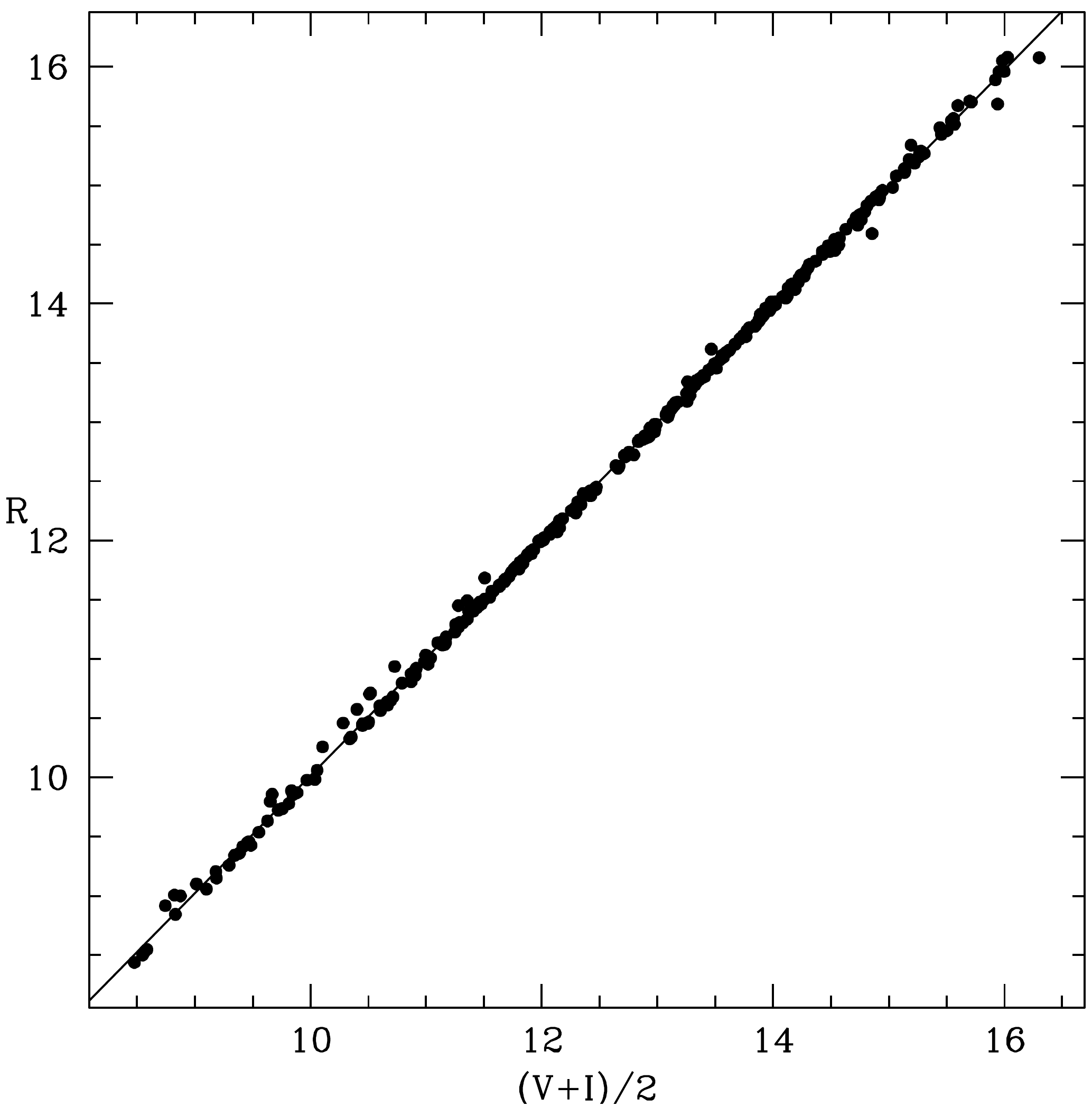}
\caption{$R$ versus $(V+I)/2$ for  the Landolt standards used to
calibrate the Johnson-Kron-Cousins photometric system
\citep{Landolt92}.}
\label{Landolt}
\end{figure}

This catalogue was calibrated on the basis of independent and
accurate $BVI$ photometry (by Stetson, private communication), already
used in \citet{Marconi14}. This photometry was transformed into the
$ugriz$ photometric system, using the equation in Table 3 in
\citet{Jordi06}, and the well defined relation between $R$ and
$(V+I)/2$ for the Landolt standards used to calibrate the
Johnson-Kron-Cousins photometric system
\citep{Landolt92}: $$R=0.003(\pm 0.001) \times (V+I)/2 + 0.08 (\pm
0.02) $$ with a $r.m.s.=0.04$. This relation is shown in
Fig. \ref{Landolt}. This means that we have calibrated photometry in
three SDSS bands ($gri$) and in four Johnson-Kron-Cousins ($BVRI$)
bands.

\begin{figure}
\includegraphics[width=0.45\textwidth]{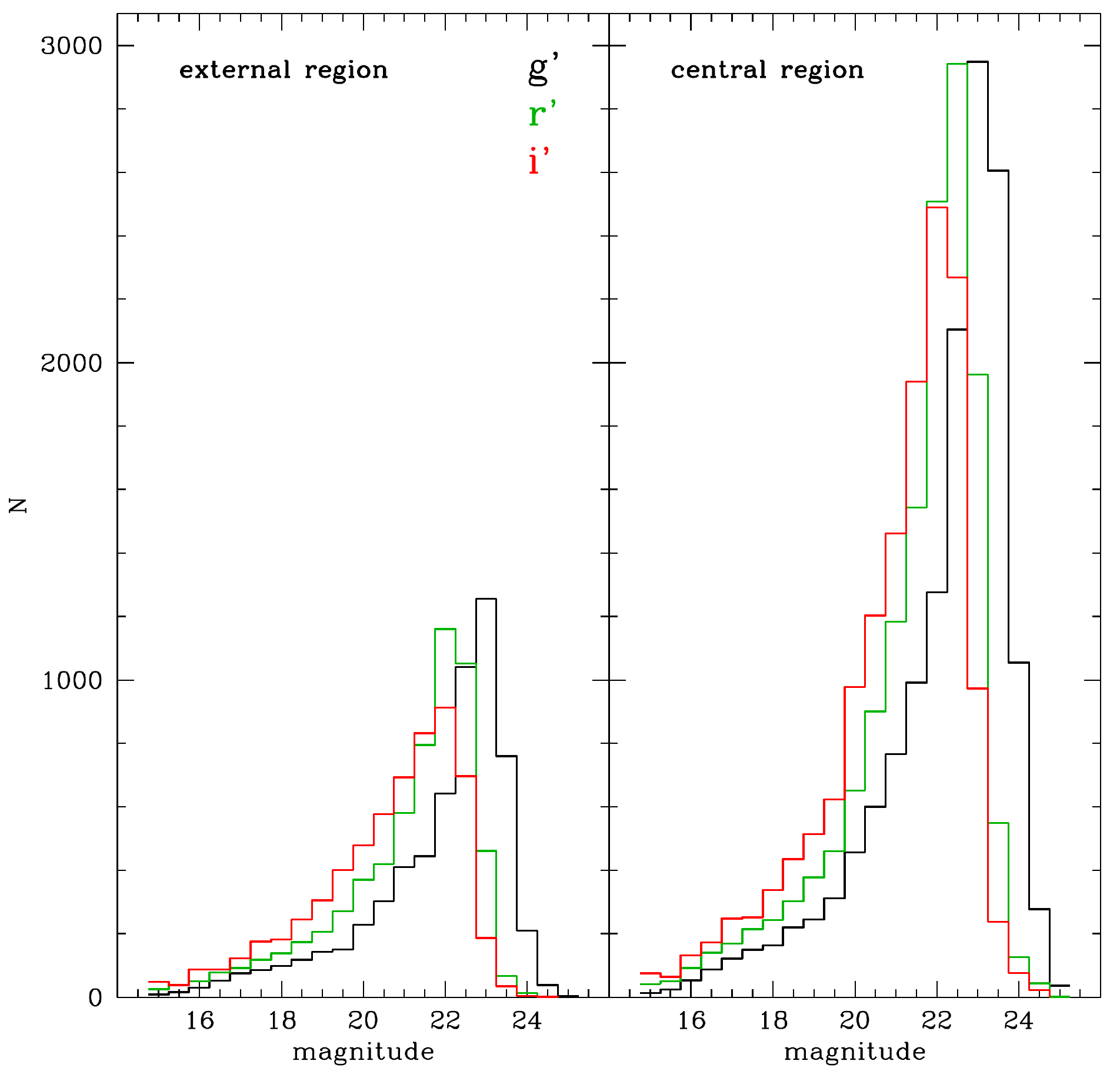}
\caption{Histogram of number of stars in magnitude
bins for the three bands, $g$ (black line), $r$ (green line) and $i$
(red line) for the external (left panel)
and central (right panel) regions.}
\label{completezza}
\end{figure}

 In our photometric data, we have areas with two exposures (the central region
overlap between F1 and F2 except for the gaps between the Omegacam
CCDs) and areas with a single observation as the quoted gaps and the
region external to the fields overlap.
In Fig. \ref{completezza}, we report the number of stars
in magnitude bins for the three bands, $g$ (black line), $r$ (green
line) and $i$ (red line) for the external (left panel) and central
(right panel) regions, respectively. From this plot we infer that to
obtain the same completeness limit for all the covered area, we need
to take $g_{\rm lim}=23$ mag, $r_{\rm lim}=i_{\rm lim}=22$ mag 
  corresponding to the limit magnitudes for single exposure areas.  The
photometric uncertainties in the $g$, $r$ and $i$ bands, within these
magnitude limits, are less than 0.05 mag.

\section{Pal~12 CMD}\label{sec-cmd}

\begin{figure}
\includegraphics[width=0.95\columnwidth]{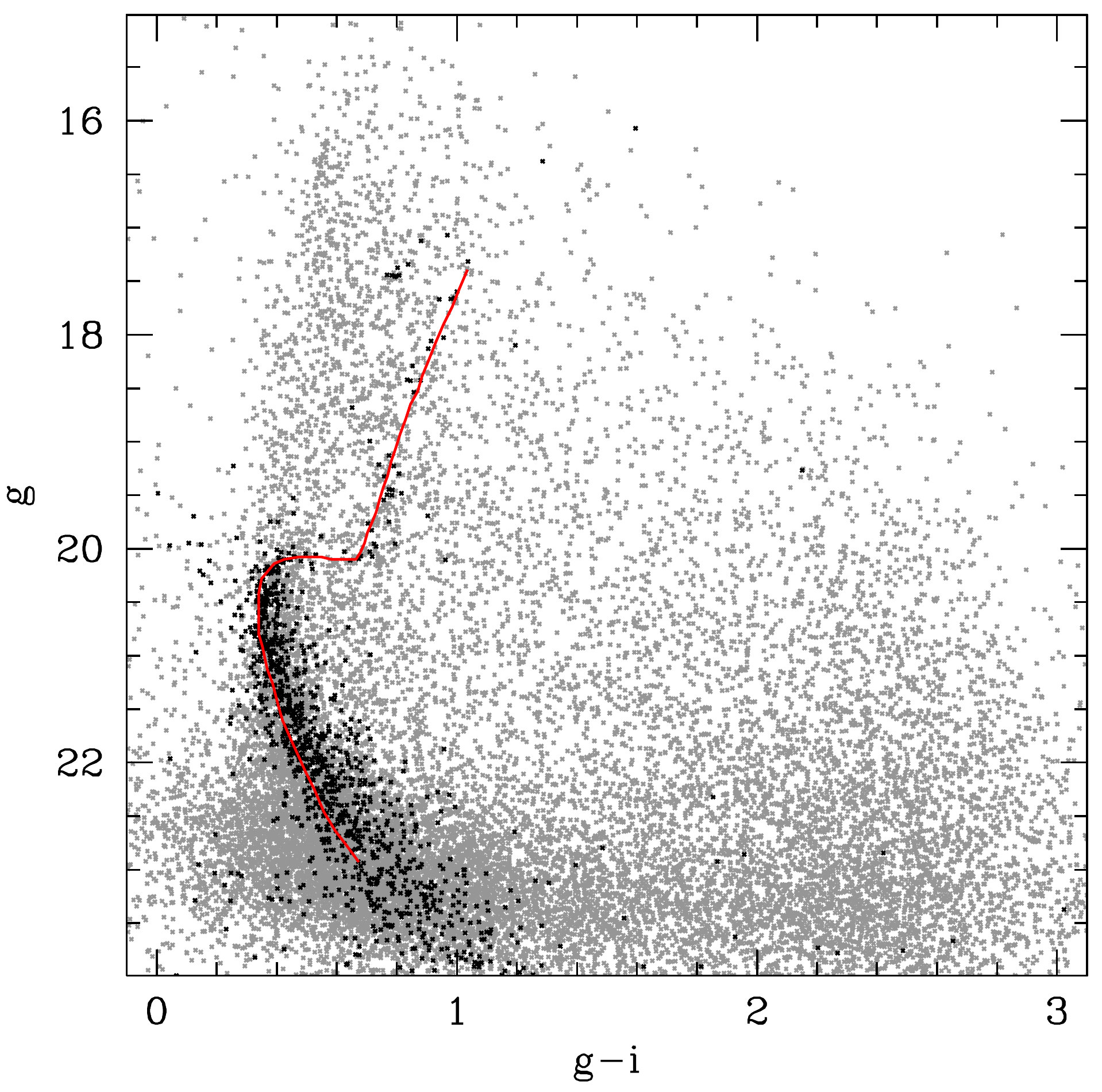}
\caption{The $g$,$g-i$ CMD for all the stars in the observed area
(grey dots). The stars in the region centered on Pal~12 and
with radius less than the half-mass radius are in black, and the derived
empirical ridge line is in red.} \label{cmd}
\end{figure}

In Fig.~\ref{cmd} we show the $g$,$g-i$ CMD for the entire area
(grey dots). Black symbols represent stars included in the region
centered on Pal~12 within the known half mass-radius
\citep[$r_h=0.03$ deg,][]{Harris96}.  The red line represents the
Pal~12 ridge line obtained taking into account the stars with
$r<0.05$ deg from cluster center.

\begin{figure}
\includegraphics[width=0.95\columnwidth]{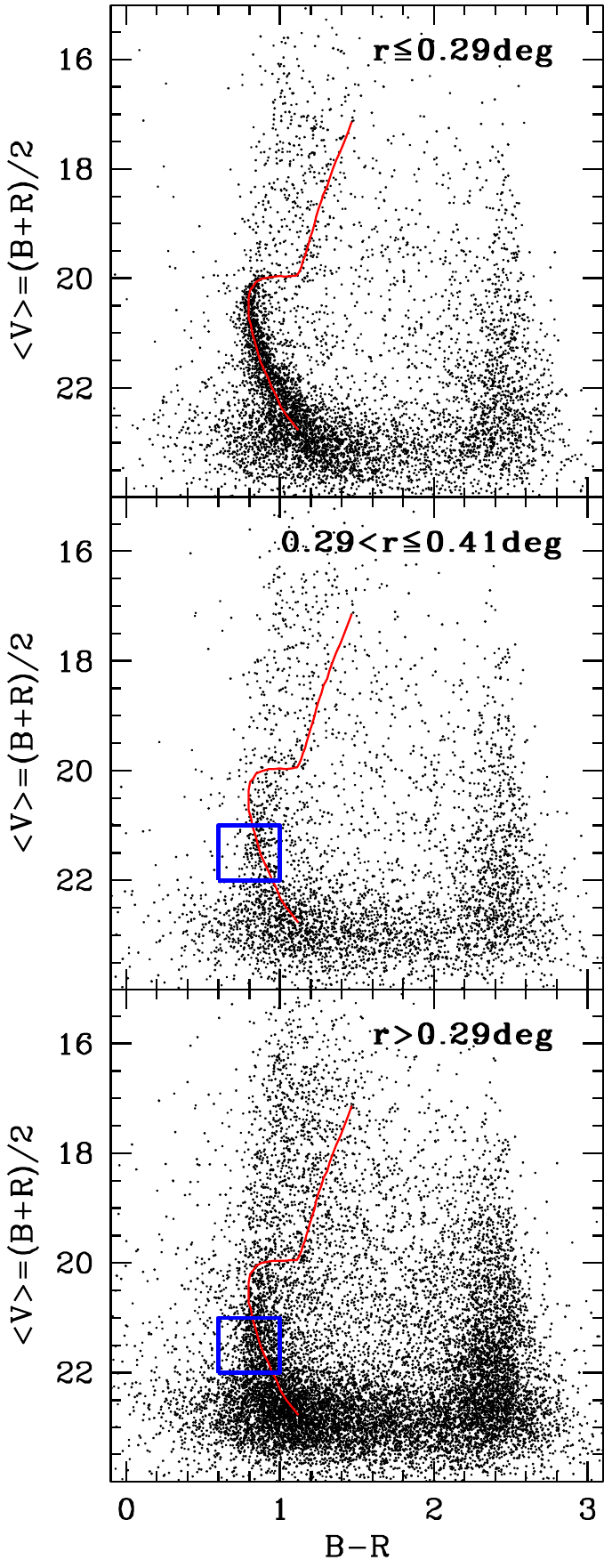}
\caption{$<V>$,$B-R$ CMDs for the stars with radius less than and
  greater than the tidal radius $r_T$ are plotted in the upper and
  lower panels, respectively. The red line is the derived empirical
  ridge line for Pal~12. The blue box shows the overdensity pointed out
by \citet{MD02}} \label{cmd_MD}
\end{figure}

In their paper (see their Fig. 1), \citet{MD02} have used a CMD with
a particular combination of bands ($<V>=(B+R)/2$ versus $B-R$) that
shows the presence of a well defined structure, distinct from the
Pal~12 main sequence, bluer than Pal~12, and overlapping the cluster
MS for $B-R\sim0.8$ mag and $V$ magnitude fainter than 21 mag.

In Fig. \ref{cmd_MD}, we show the CMD for stars with $r\leq r_T$
(upper panel) and $r>r_T$ (lower panel). In the middle panel, we plot
the CMD for the annular region with $r>r_T$ and area equal to that
included within one tidal radius ($0.29 < r \le 0.41$ deg).  The red line
represents the Pal~12 ridge line as in Fig. \ref{cmd}. To better
  distinguish the stellar overdensity identified by \citet{MD02} 
  for $<V>$ fainter than about 21 mag and $B-R$ color
between 0.6 and 1.0 mag, we use a
  blue box to mark this CMD region. The presence of this overdensity
  in our data will be more evident using the luminosity functions
  (LFs) in the next section.

Using a control field located in the north at $l=28.7$
and $b=42.2$ deg, \citet{MD02} showed that this structure cannot be
ascribed to the Galactic halo. They pointed out that this population shows a  significant
width in color, signature of a possible range in metallicity and/or
age and/or depth, typical of a stream of a dwarf galaxy. Moreover,
they conclude that on the
basis of its position and extension on the sky, this structure could
be part of the Sgr tidal stream \citep[see e.g.][]{Newberg02}.

To analyse our data and get information about the stellar content in Pal 12 and in the above mentioned
overdensity (see next section), we used the stellar population
synthesis code SPoT (Stellar POpulation Tool,
\citealt{Brocato99,Brocato00,Raimondo05,Raimondo09})  to compute both
synthetic CMDs and LFs. Here, we used the
version of the code optimized to reproduce the properties of poorly
populated stellar systems, like those studied in the present work.  For
this purpose, the code relies on Monte Carlo techniques to
generate stars according to an initial mass function (IMF), in a
way that takes into account statistical effects due to the number of
stars in the different evolutionary phases in low-luminosity stellar
systems. In the present version the IMF is assumed to be a Kroupa-like
function from 0.1 to 100 M$_\odot $, the evolutionary tracks are from
the Basti database \citep[][and references therein]{Pietrinferni04},
and the stellar atmospheres come from \citet{Castelli03}. 

\begin{figure}
\centering
\includegraphics[angle=0, trim= 0 6.5cm 0 4cm, clip=true, width=1\columnwidth]{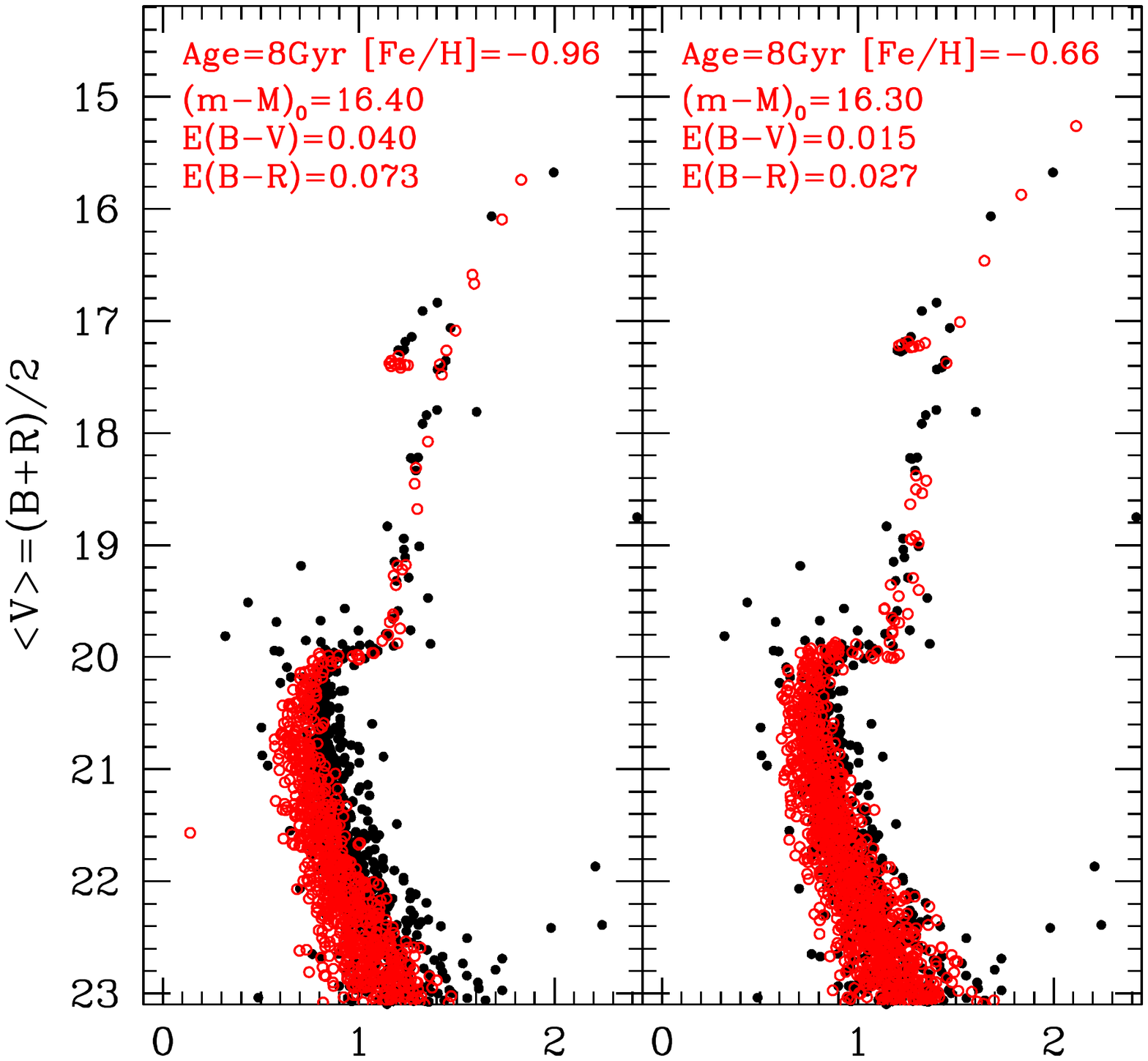}
\includegraphics[angle=0, trim= 0 6.5cm 0 4cm, clip=true, width=1\columnwidth]{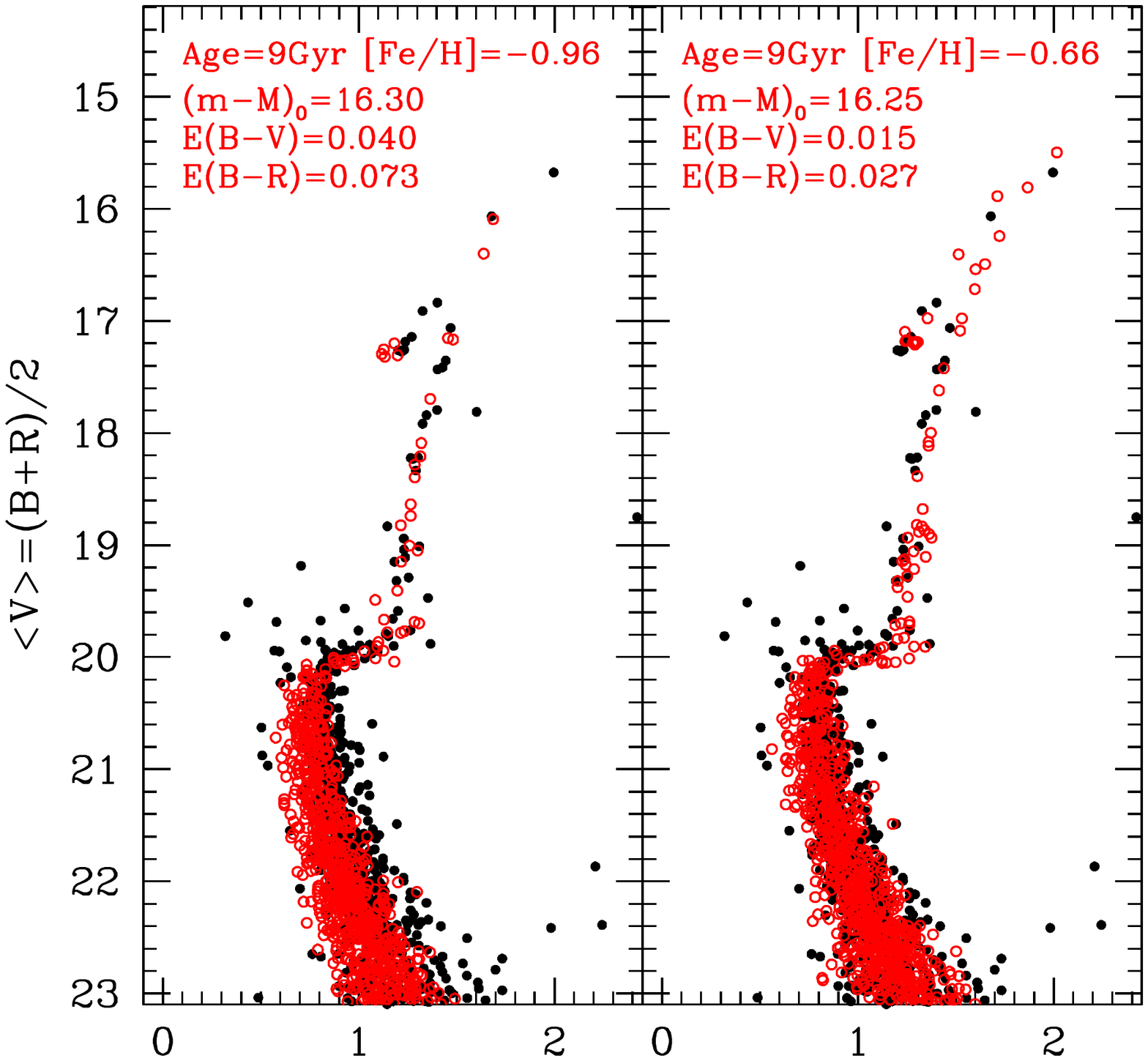}
\includegraphics[angle=0, trim= 0 5cm 0 4cm, clip=true, width=1\columnwidth]{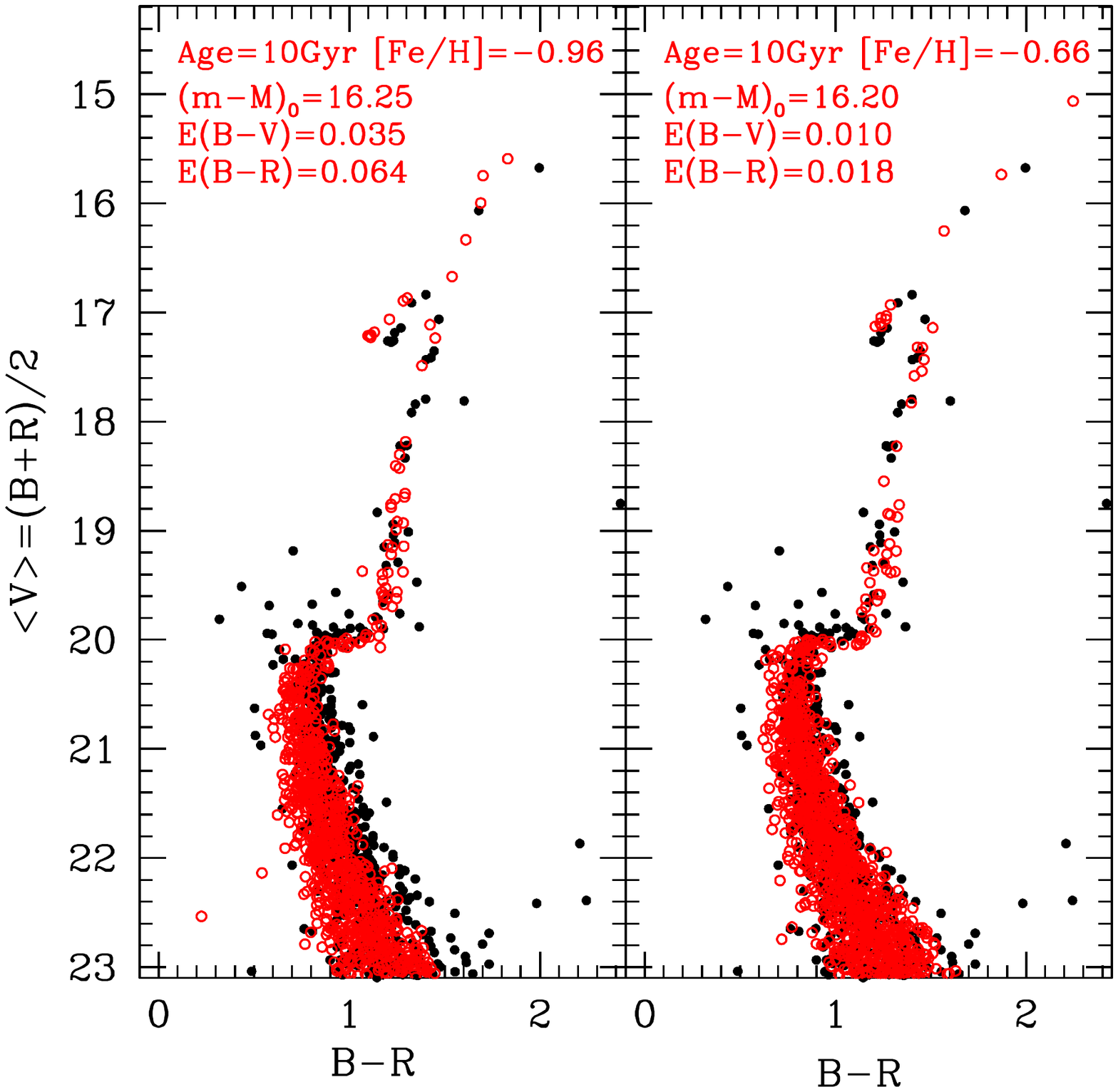}
\caption{Three simple stellar populations (red open circles) aged 8
(top), 9 (middle) and 10 Gyr (low) with $[Fe/H]=-0.96$~dex (left) and
$[Fe/H]=-0.66$~dex (right) are superimposed on the CMD of stars within the
half-light radius of Pal~12 (black filled circles). Each simulation
accounts for a total V magnitude of $\sim -3.7$ mag, corresponding to
the magnitude at the half-light radius as derived from Harris's
catalogue. Some photometric uncertainty is also included. The fitting
values for distance modulus and interstellar reddening are labeled in
figure. The extinction law adopted is that by \citet{Cardelli89}.}
\label{cmdhalf}
\end{figure}

Earlier studies of Pal~12 CMD (see Sect.~1) suggested that the cluster
is somewhat younger than the vast majority of Galactic GCs, with a
metallicity of $[Fe/H]\sim-1.0$~dex
\citep[e.g.][]{GO88,Stetson89,Rosenberg98}. Spectroscopic studies of
a few cluster stars concluded that Pal 12 is quite metal-rich, with
values ranging from $[Fe/H] = -0.6$ dex  \citep[][from low-resolution
spectroscopy]{Armandroff91} to $[Fe/H] = -1.0$ \citep{Brown97} and
$[Fe/H] \sim -0.8$~dex \citep{Cohen04} from high-resolution
spectroscopy.  Interestingly, these studies found that Pal~12 stars do
not show $\alpha$-elements enhancement, but rather very peculiar abundance
ratios \citep{Cohen04}, which seem to further support the scenario of
a cluster tidally stripped from the Sgr galaxy
\citep{Cohen04,Sbordone07}. Bearing in mind this complex picture, in
Fig.~\ref{cmdhalf} we plot the CMD of the stars within the half-light
radius of Pal~12 ($r\leq r_{h}$)---in such a way as to avoid large
field star contamination---together with three synthetic simple
stellar populations with ages t=8, 9, and 10$\,$Gyr, metallicity
$[Fe/H]=-0.96$~dex (left panels), and $[Fe/H]=-0.66$~dex (right
panels), both with a solar-scaled composition ([$\alpha$/Fe]=0.0).
The fitting values for the distance modulus and interstellar reddening
(as labelled in Fig. \ref{cmdhalf}) depend on the adopted age and
metallicity. Using the values labeled in Fig. \ref{cmdhalf}, the
inferred mean distance and reddening are $(m-M)_0=16.3\pm 0.1$ mag,
which is  consistent---within the uncertainties---with estimates
published in the literature \citep[e.g.][and references
therein]{Rosenberg98} and $E(B-V)=0.03 \pm 0.01$ mag in agreement with
$E(B-V)$ by \citet{Schlegel98} (recalibrated by \citet{Schlafly11},
see also Fig. 8 in \citet{Marconi14}).  On this basis, we confirm an
age of $8 - 10$ Gyr, and a rather high metallicity ($[Fe/H] -0.96$ to
$-0.66$ dex) for a GC in the Galaxy's outer halo. Note that a small
contamination by stars belonging to the Sgr stream is not excluded.

\section{Luminosity functions}\label{sec-lum}

\begin{figure}
\centering
\includegraphics[angle=0, trim= 0 5cm 0 4cm,clip=true, width=1.05\columnwidth]{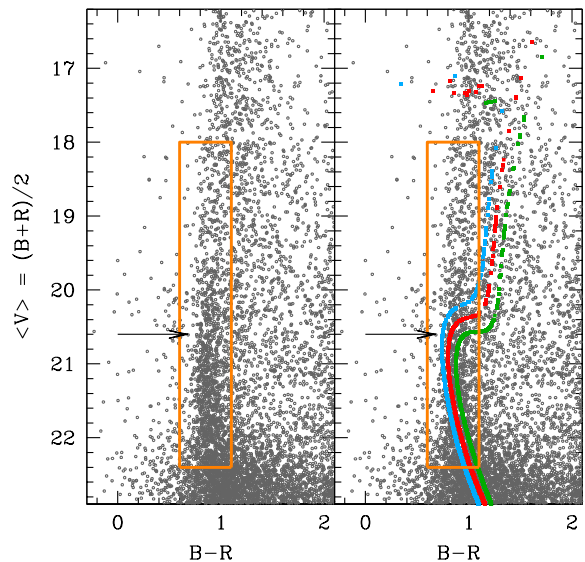}
\caption{Left panel: CMD of stars in the extra-tidal region of F1
  field (grey circles). The selection box is shown in orange.
  Right panel: Synthetic stellar populations (squares) with t=13~Gyr and $[Fe/H]=-1.27$~dex
  (sky blue), $[Fe/H]=-0.96$~dex (red), and $[Fe/H]=-0.66$~dex (green), representative of the old
  components in the Sgr dSph galaxy, are
  overplotted. A distance modulus of 16.30 mag and $E(B-V)=0.03$ mag are
  adopted.  In each panel the arrow indicates the feature at
  $<$$V$$>=20.6$\,mag, see text. }
\label{streampop}
\end{figure}

In their paper, \citet{MD02} compared the luminosity function (LF) of
the apparent stellar overdensity to that of the control field, and to a
theoretical LF obtained in the same color and magnitude range and with
the same distance and reddening as Pal~12. From this comparison, the
authors argued the presence of a stellar population with an age of
12.6 Gyr and $Z=0.001$, although the number of detected stars is too
small to reach a firm conclusion. Our analysis is based on a wider FOV
that contains a larger number of stars. Therefore, we proceeded to
investigate the stellar overabundance shown in Fig.~\ref{cmd_MD} by
analysing the LF of stars in the extra-tidal region, where the
contribution of the Pal~12 stars is negligible. For this purpose we
take into consideration only the F1 field and select stars with
$r>r_{T}$ in the $<V>$ magnitude range 18--22.4 mag and $(B-R)$ color
between 0.6 and 1.1 mag \citep[as in][]{MD02}. $V=22.4$ mag is
  the completeness limit in the visual band on the basis of the $g$,
  $r,$ and $i$ limit
  magnitudes obtained in Section \ref{sec-red}. These choices allow us
not only to avoid objects with large photometric uncertainties and serious field
contamination, but also to take into consideration the possible presence of
multiple stellar populations, and/or a distance spread. The left panel of
Fig.~\ref{streampop} illustrates the CMD of all extra-tidal stars in
F1 (grey circles) and the color-magnitude box (orange rectangle)
considered in our analysis.

\begin{figure}
\centering
\includegraphics[angle=0, trim= 1cm 12.2cm 1cm 4cm, clip=true, width=\columnwidth]{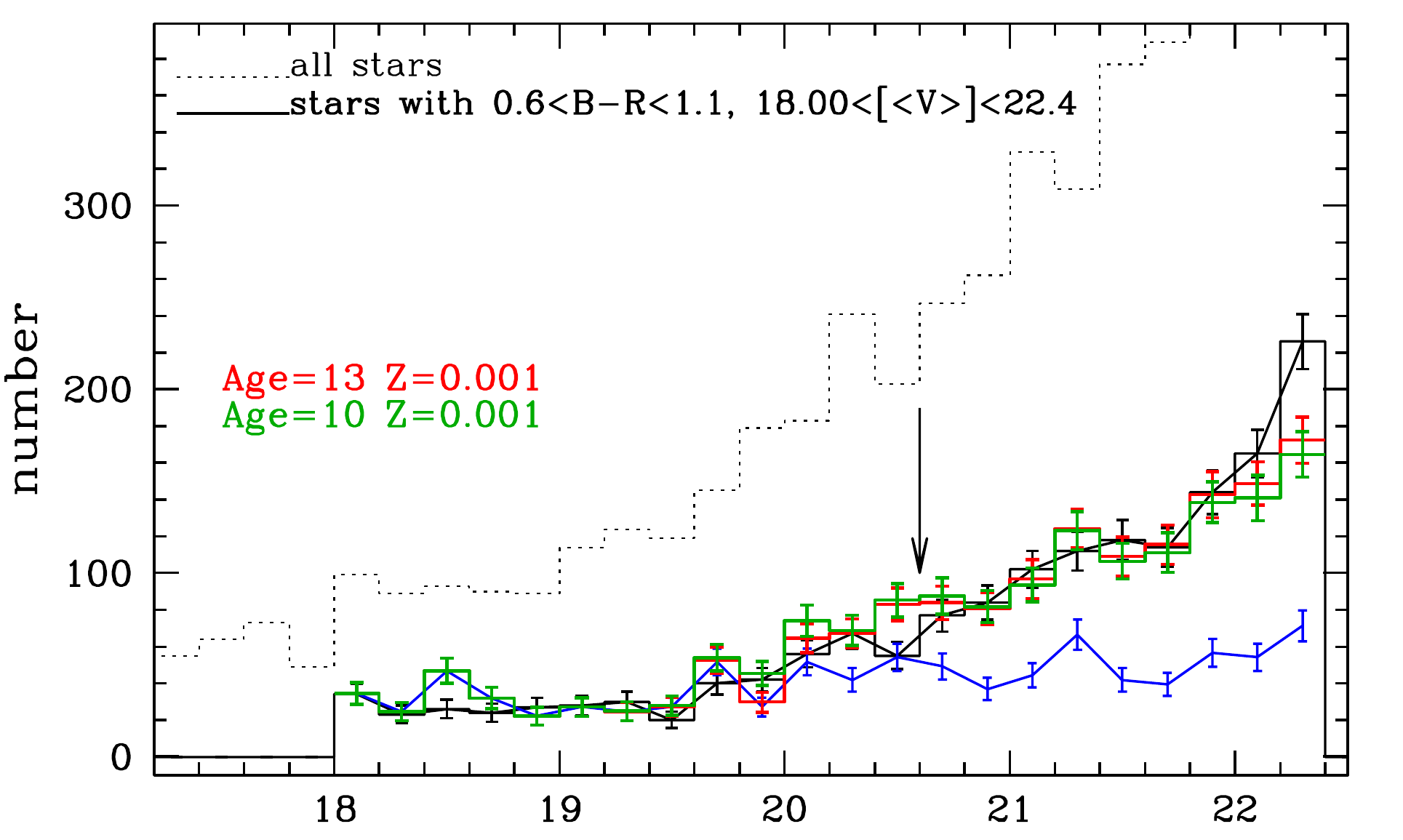}
\includegraphics[angle=0, trim= 1cm 12.2cm 1cm 4cm, clip=true, width=\columnwidth]{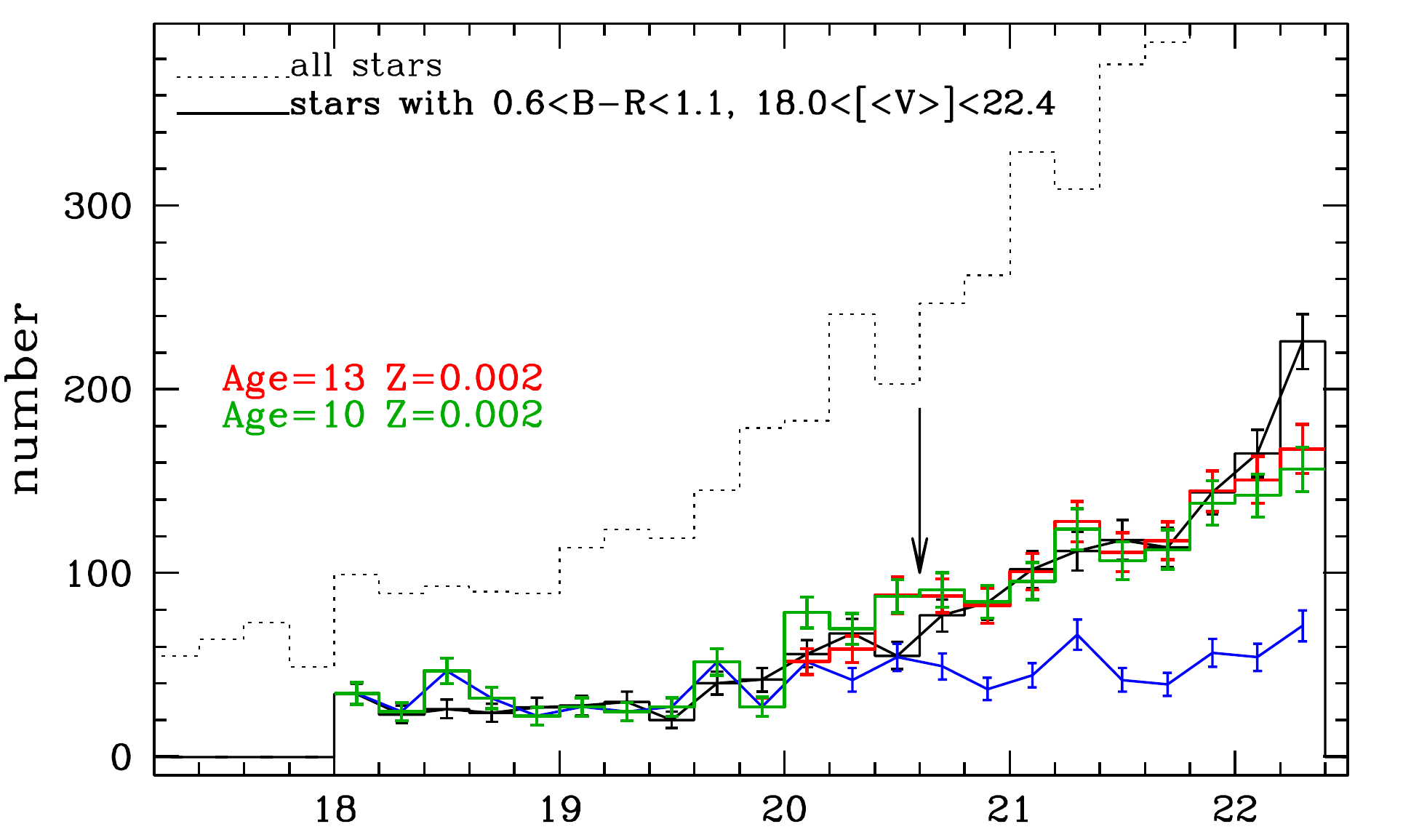}
\includegraphics[angle=0, trim= 1cm 11cm   1cm 4cm, clip=true, width=\columnwidth]{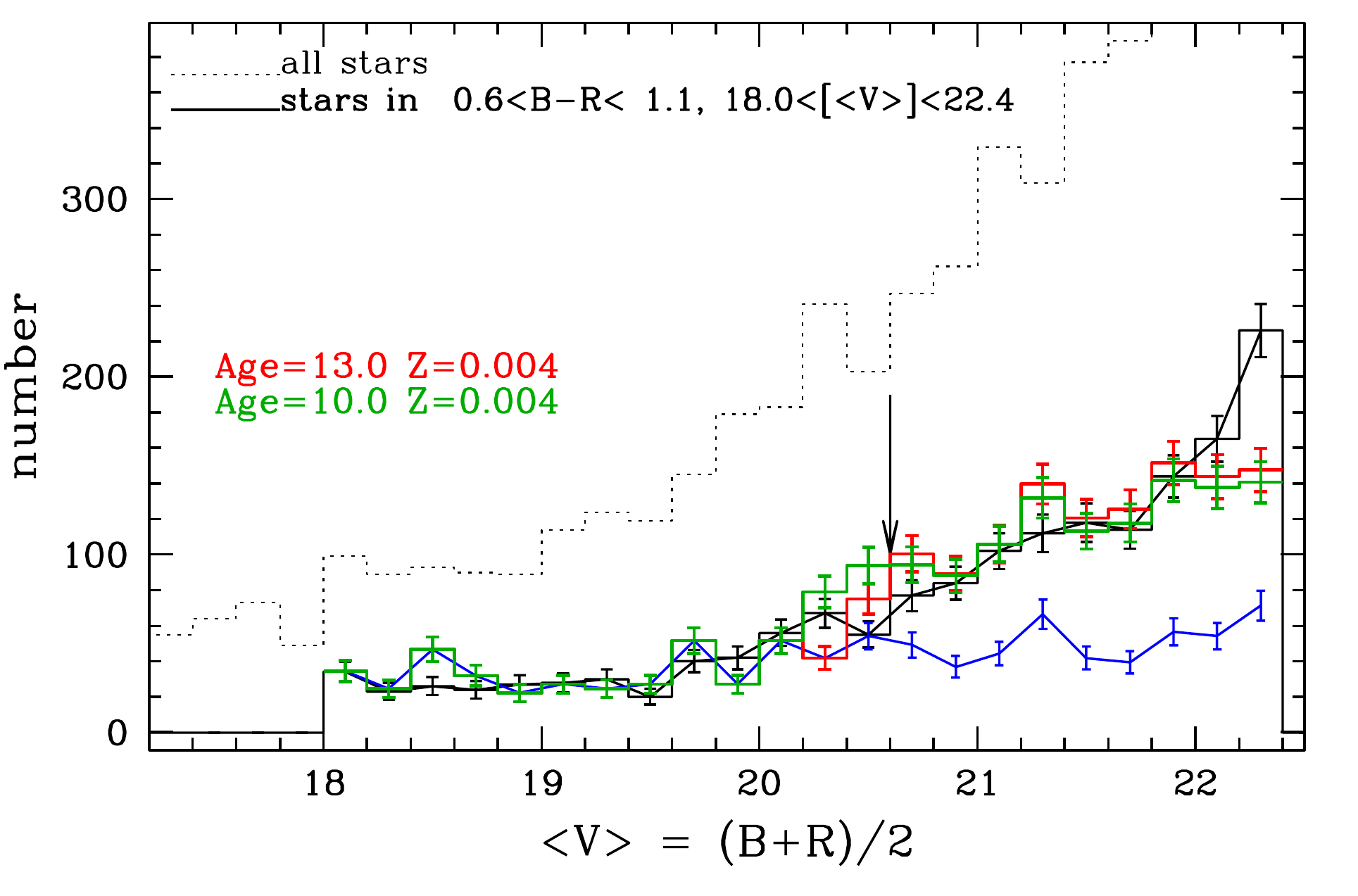}
\caption{Luminosity functions of all extra-tidal-field stars in F1
(black dotted line) and of stars in the selection box (black solid
line) are plotted together with the LF of stars in the control field
(blue line, see text for details). From upper to lower panel: the
observed LFs are compared with the synthetic ones obtained for stellar
populations with the indicated age and metallicity, for which a
distance modulus of 16.3 mag and $E(B-V)=0.03$ mag have been
adopted. The arrow is drawn at magnitude $<$$V$$>=20.6$\,mag (see text
and Fig.~\ref{streampop}). }
\label{lfs1}
\end{figure}

In each panel of Fig.~\ref{lfs1} we show the LF of stars in the
selection box (black solid line), compared to that of all stars in the
extra-tidal field (black dotted line). The former LF likely includes
stars belonging to the Sgr stream, with a contamination of Galactic
field and possibly of extra-tidal Pal 12 stars. As stated above, it
has been proved with many different tracers that the disruption of the
Sgr dSph galaxy produces huge tidal tails extending for many tens of
kiloparsecs from the parent galaxy, with tidally stripped stars
wrapping a full 360$^{\circ}$ around the celestial sphere \citep[see,
e.g.][]{Newberg02,Majewski03,MD04,Belokurov14}. Recent simulations
also predict the existence of several arms extending to hundreds of
kiloparsecs \citep[see e.g.][]{Dierickx17}. Since we did not observe a
reference field far away from the 
stream, to investigate the star overdensity we considered the
contribution of Galactic field stars using the LF in the control
field, observed by \citet[][see their Fig. 2]{MD02} and rescaled to
the area covered by our extra-tidal region (blue solid line in
Fig. \ref{lfs1}).  From Fig.~\ref{lfs1} one may note that in the
magnitude range from $<V>\,\sim$\,18 to $\sim$\,19.9\,mag the
control-field and the extra-tidal star LFs are comparable. Hence, in
this magnitude range, we do expect stars likely belonging to the
Galaxy only, while for magnitudes fainter than 20.6 mag (this value is
marked with an arrow in Fig.~\ref{streampop}) the observed LF displays
an overabundance which increases with the magnitude, and likely
includes stars of the Sgr stream.

\begin{figure}
\centering
\includegraphics[angle=0, trim= 1cm 11cm   1cm 4cm, clip=true, width=\columnwidth]{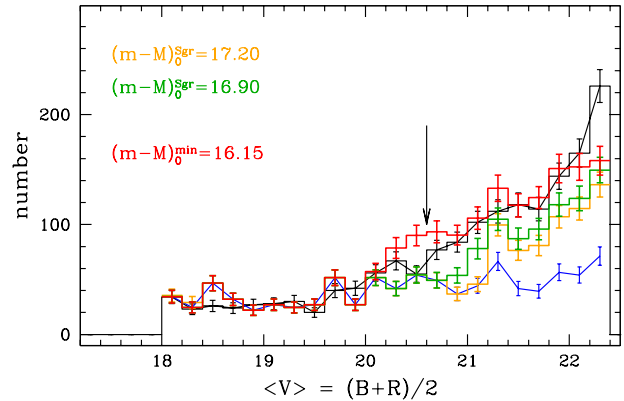}
\caption{The LF of the extra-tidal stars in the selection box
(black) is compared to the LF of a synthetic population with age 13$\,$Gyr
and $[Fe/H]=-0.96$~dex located at a minimum distance of
$(m-M)_0=16.15$~mag (red) and at the distance of Sgr galaxy, for which
we consider a minimum (maximum) value of $(m-M)_0=16.9$~mag
(17.2~mag), green and orange line, respectively.}
\label{lfs2}
\end{figure}

To tentatively analyse the stellar content of such a stellar overdensity,
we consider the limiting case of a negligible amount of
extra-tidal Pal~12 stars (i.e., none).  According to  \cite{Bellazzini06a}, the
bulk of the stellar population in the Sgr galaxy is mainly composed of
stars older than $\sim$8\,Gyr with metallicity Z$\la$0.004, with the
age limit decreasing if the metallicity increases to $Z=0.008$. Among
the population older than 1--2 Gyr, \citet{Monaco03} established from the analysis of a sample of
horizontal Branch stars that a
fraction around $\sim$10\% have an age $\ga$~10~Gyr and
metallicity $[Fe/H] \la -1.3$~dex. Additional populations composed of
very metal-rich stars with an age of 1$\,$Gyr or younger have been
detected, e.g., by \citet{Bonifacio04}.  Moreover, high-resolution
spectroscopic studies showed that some abundance anomalies are present
also in stars of the Sgr dwarf spheroidal galaxy
\citep[e.g.][]{Carretta10,McWilliam13}. Anyway, taking into account
the photometric and other uncertainties,  the assumption of
solar-scaled models to predict theoretical LFs for the stream stars
does not introduce biases into our analysis.  Assuming that the
overdensity is located at nearly the same average distance adopted for
Pal~12 [$(m-M)_0=16.3$ mag with an uncertainty of $\sim$0.1 mag], with
an interstellar reddening of $E(B-V)=0.03$ mag, we perform a
comparison with several  synthetic populations with different ages and
metallicities. From this analysis, the feature at 20.6 mag (showed in
the right panel of Fig. \ref{streampop}) appears to be consistent with
old ($t\ga 10$ Gyr) populations with  the best-fit synthetic
population model corresponding to $t\sim$13~Gyr and mean metallicity
Z$\la$0.004 with the more metal-poor stars at bluer colors and the
more metal-rich stars at redder colors. This result is shown in the
right panel of Fig. \ref{streampop} in which we plot the synthetic
populations with  $t=13$ Gyr and metallicities $Z=$0.001, 0.002, 0.004
(corresponding to $[Fe/H]= -1.27$, $-0.96$, and $-0.66$~dex).

In the three panels of Fig. \ref{lfs1}, we also show the theoretical
LFs resulting from the sum of the control-field stars and of the
synthetic stars with  $t$=10 and 13 Gyr and $Z=$0.001, 0.002, 0.004,
always assuming a Pal 12 distance of 16.3 mag.  In detail, for each
assumed age and metallicity, the synthetic LF is derived by averaging
the number of stars in each magnitude bin over 100 CMD simulations,
with uncertainty due to statistical effects. A variation of the Pal~12
distance modulus from 16.2 to 16.4 mag does not cause visible
changes. The number of stars in each simulation is chosen to be equal
to the observed number of stars in the selection box, after
subtracting control-field stars. This means about 690 stars, likely
belonging to the stream, and distributed at the distance of Pal 12. As
already stated, we assumed a negligible number of Pal~12 stars in the
extra-tidal field, so that the previous number represents an upper
limit for stream stars. Each of the adopted models can produce an
equally good fit taking into account the uncertainties due to
photometric measurements, interstellar reddening, and distance.  On
this basis, we cannot obtain firm constraints on the age and chemical
composition of the overdensity, likely populated by a mixture of old
and metal-poor/intermediate-metallicity populations.  This hypothesis
is in agreement with \citet{Bellazzini06b} who, analysing the
horizontal branch stars in the core of the Sgr galaxy and in a wide
field located tens of kpc away from the center of the galaxy itself,
concluded that old and metal-poor populations appear to be
preferentially stripped from the Sgr galaxy during the past
peri-Galactic passages with respect to the intermediate-age
intermediate-metallicity population that presently dominates its bound
core.

To investigate the impact of a possible distance spread of stream
stars, we considered different assumptions about their distance
modulus, from a value as near as $(m-M)_0=$16.15~mag  (d$\sim$17~kpc),
suggested by \citet{MD02}, up to the Sgr distance. Note that the lower
limit is within the uncertainty of the distance for Pal 12 published
in the literature \citep[e.g.][]{GO88}: $(m-M)_0=16.1\div16.5$~mag.
On the other hand, the available estimates of the Sgr distance range
from a short value of $(m-M)_0=16.90\pm0.15$~mag \citep{Alard96} to a
long distance of $(m-M)_0=17.25^{+0.10}_{-0.20}$~mag
\citep{Bellazzini99}. More recent estimates have been provided with
different methods, and, in general, appear to favor long
distances. For instance, from the analysis of the red giant tip
\citet{Monaco04} obtained $(m-M)_0 = 17.10 \pm 0.15$~mag (d=$26.3 \pm
1.8$~kpc) while from a sample of Sgr RR Lyrae stars of the MACHO
database, \citet{KC09} found $d=24.8$~kpc [$(m-M)_0 \sim
16.97$~mag]. More recently from the analysis of a large RR Lyrae stars
sample from the OGLE-IV observations, \citet{Ogle16} obtained for the
cluster M54 in the core of the galaxy
$d=26.7\pm0.03_{stat}\pm1.3_{sys}$~kpc [$(m-M)_0 \sim 17.13$~mag]. In
order to take into account these differences, we recomputed the
theoretical LF assuming distances spanning from the shortest to the
longest values mentioned above. In Fig. \ref{lfs2} we compare the LF
of the F1 extra-tidal stars in the selection box (black solid line) to
the synthetic LF obtained from a population with age 13 Gyr, and
$[Fe/H]=-0.96$ dex located at 17 kpc (red line), 24 kpc (green line,
the shortest distance value of the dwarf galaxy), and 27.5 kpc (orange
line, the longest distance). The figure suggests that, even if a
better agreement is found for a Pal 12 distance modulus of
$(m-M)_0=$16.3 mag (Fig. \ref{lfs1}), we cannot exclude the presence
of a fraction of stars with longer or slightly shorter distances.

\section{Star counts around Pal~12}\label{sec-counts}

\begin{figure}
\includegraphics[width=0.95\columnwidth]{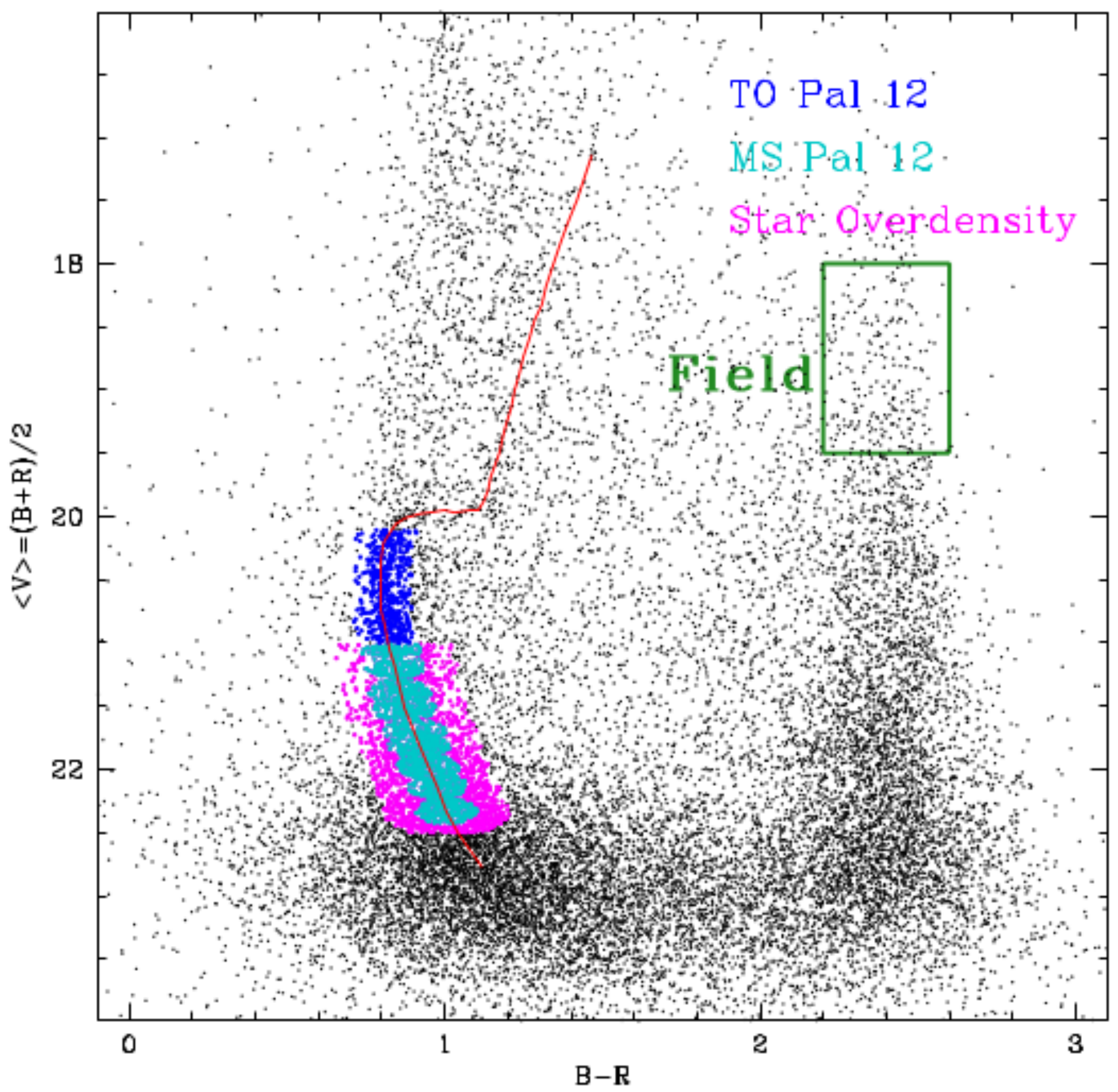}
\caption{The $<V>$,$B-R$ CMD with the stars with a
distance of $\pm 0.1$ mag from the ridge line (the red line) highlighted: blue and
cyan for the Pal~12 TO and MS stars, respectively. The magenta stars
have a color distance ranging between 0.1 and 0.2 mag from the ridge
line. The magenta and cyan regions include the star overdensity
identified in Fig. \ref{cmd_MD}.}\label{cmd_MD_col}
\end{figure}

On the basis of the CMD for the stars included in the tidal radius
(see top panel of Fig. \ref{cmd_MD}) and following the procedure already
adopted in \citet{Marconi14}, we consider as possible cluster members
the stars lying in a range of $\pm 0.1$ mag around the ridge line of
Pal~12, marked in blue and cyan in Fig. \ref{cmd_MD_col}. Blue dots
represent the stars apparently belonging to the Pal~12 turnoff
(TO); this CMD region is identified as ``TO Pal~12''. The cyan CMD
region includes stars possibly belonging to the Pal~12 MS, and
therefore identified as ``MS Pal~12''. Magenta stars are those with a color
distance from the ridge line ranging between 0.1 and 0.2 mag. Cyan and
magenta regions include Pal~12 MS stars but also stars belonging to
the star overdensity already discussed. Finally, the region included in the
green rectangle, labelled as {\it Field}, is not expected to host
cluster members and can be used for comparison in our analysis. We
adopt different {\it Field} rectangle selections always obtaining the
same result within the Poisson errors (see below).

\begin{figure}
\includegraphics[width=0.95\columnwidth]{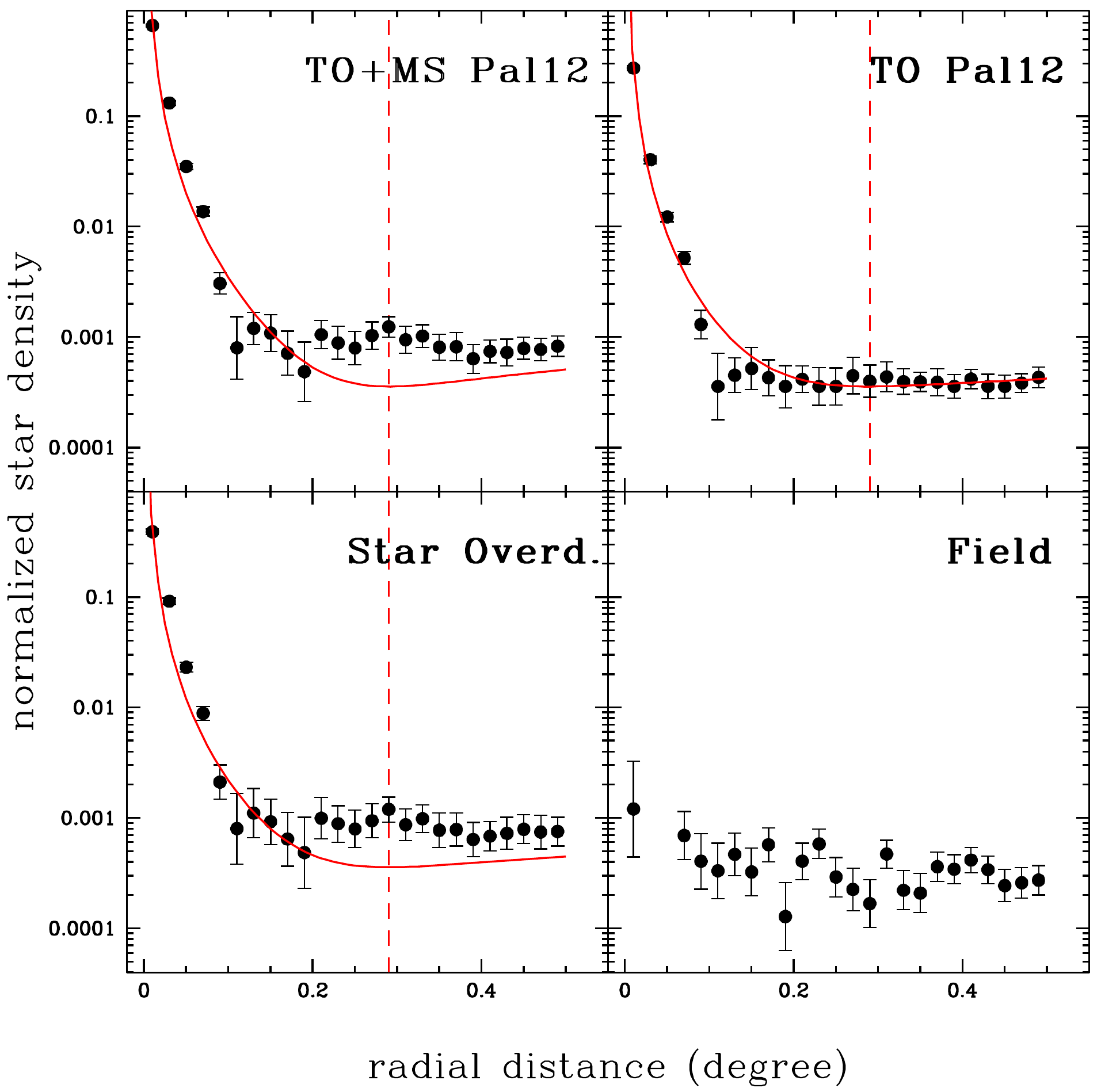}
\caption{Radial counts of the normalised star densities for  different
star selections in Fig. \ref{cmd_MD_col} (see text for details). The
red solid line represents the King profile of Pal~12 for the core radius
and tidal radius by \citet{Harris96} and the red  vertical dashed line
marks the $r_T$ value. }\label{cont}
\end{figure}

In Fig. \ref{cont}, we plot the radial counts of the normalised star
densities for the different star selections in
Fig. \ref{cmd_MD_col}\footnote{to plot these counts and the King
profile in a logarithmic scale we add to the radial counts the average
of the ``Fields'' counts.}: i) in the upper-left panel, we show
the blue and cyan stars, labelled as ``TO $+$ MS Pal~12'', ii) in the
upper-right panel are the results obtained using the blue stars, labelled
as ``TO Pal~12'', iii) in the bottom-left panel, we show the results
for the cyan and magenta stars, labelled as ``Star Overdensity'',  and
iv)  in the bottom-right panel are the star counts within the {\it Field}
green rectangle. To obtain these counts, we adopted 25 equally spaced
annuli (1.2 arcmin), from the centre up to 0.5 deg, corresponding to
about two tidal radii. For each selection, the star density in the $i$-th
annulus is normalised to the total star density computed in the circle
with the radius of 0.5 deg.  The error bars result from the
propagation of the Poisson errors, and the vertical dashed line
indicates the nominal Pal~12 tidal radius of 0.29 deg
\citep{Harris96}. For all the  selections, we obtain the typical
profile of the globular cluster \citep[see e.g.][]{King66,Wilson75},
except, as expected, for the {\it Field} stars. The red line
represents the King profile for the nominal values of Pal~12 core
radius and tidal radius. For the region around the Pal~12 TO, the star
count distribution follows the computed King profile. In the other two
regions (including the stellar overdensity), the normalised stellar density
shows a shift toward higher values for radii larger than 0.2
deg. This shift is anomalous and is probably due to the presence
of the overdensity. This might affect the determination of the King
profile parameters, but it could also be due to the procedure adopted
to select Pal~12 stars. To check this behaviour,  we use an innovative
and more complex approach to perform a more accurate separation of
field and cluster stars. In particular, we adopt a procedure  similar
to that  suggested by \citet{DiCecco15} and \citet{Calamida17} that we
call a ``3D procedure'' to distinguish it from the previous one.

\begin{figure}
\includegraphics[width=0.95\columnwidth]{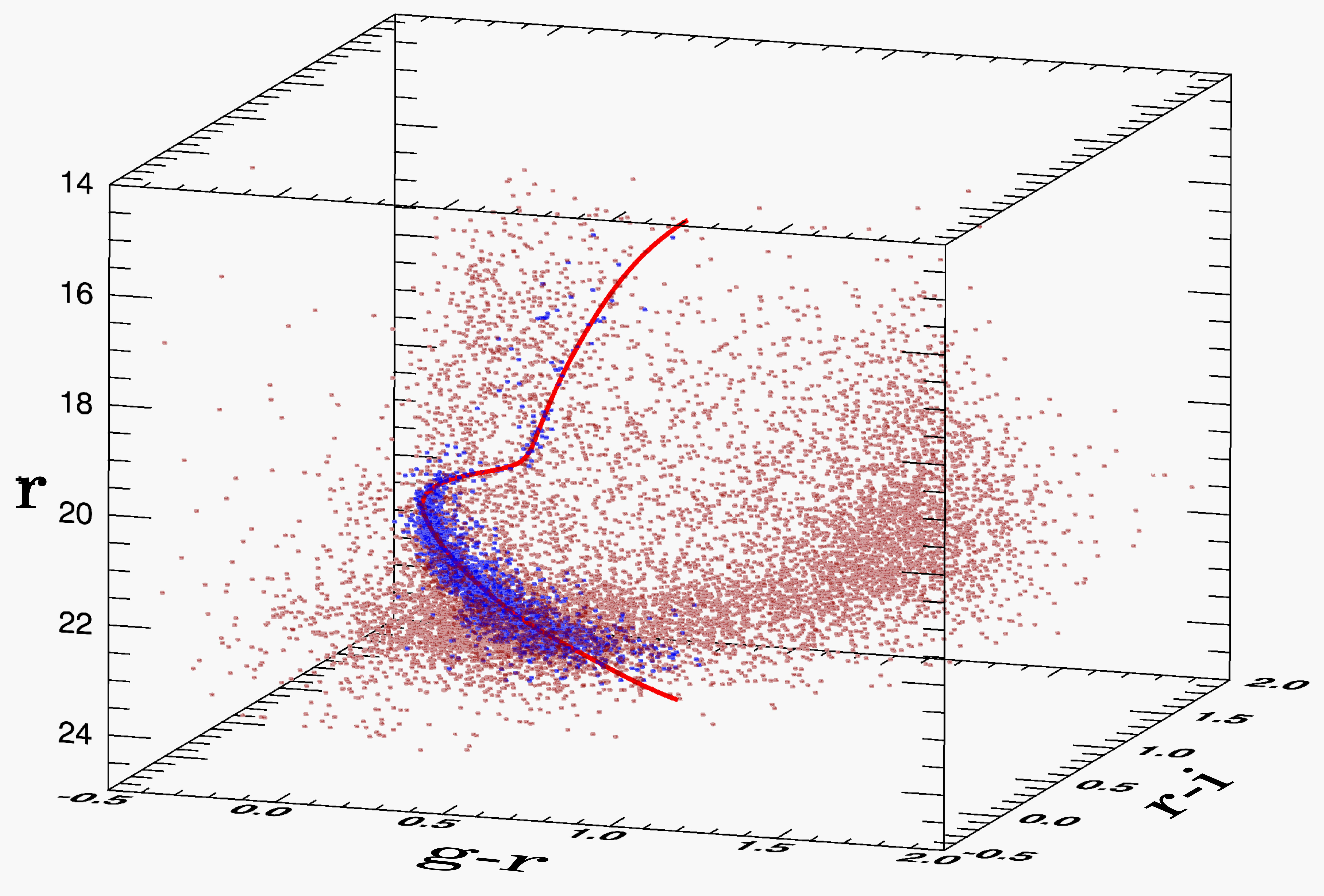}
\caption{$r$, $g-r$, $r-i$ CMD of Pal~12. The blue dots are candidate cluster stars, while the pink
dots are candidate field stars. The red solid line shows the 3D ridge line. See
text for more details concerning the selection criteria based on the
``3D procedure''. }\label{rgrgi}
\end{figure}

\begin{figure}
\includegraphics[width=0.95\columnwidth]{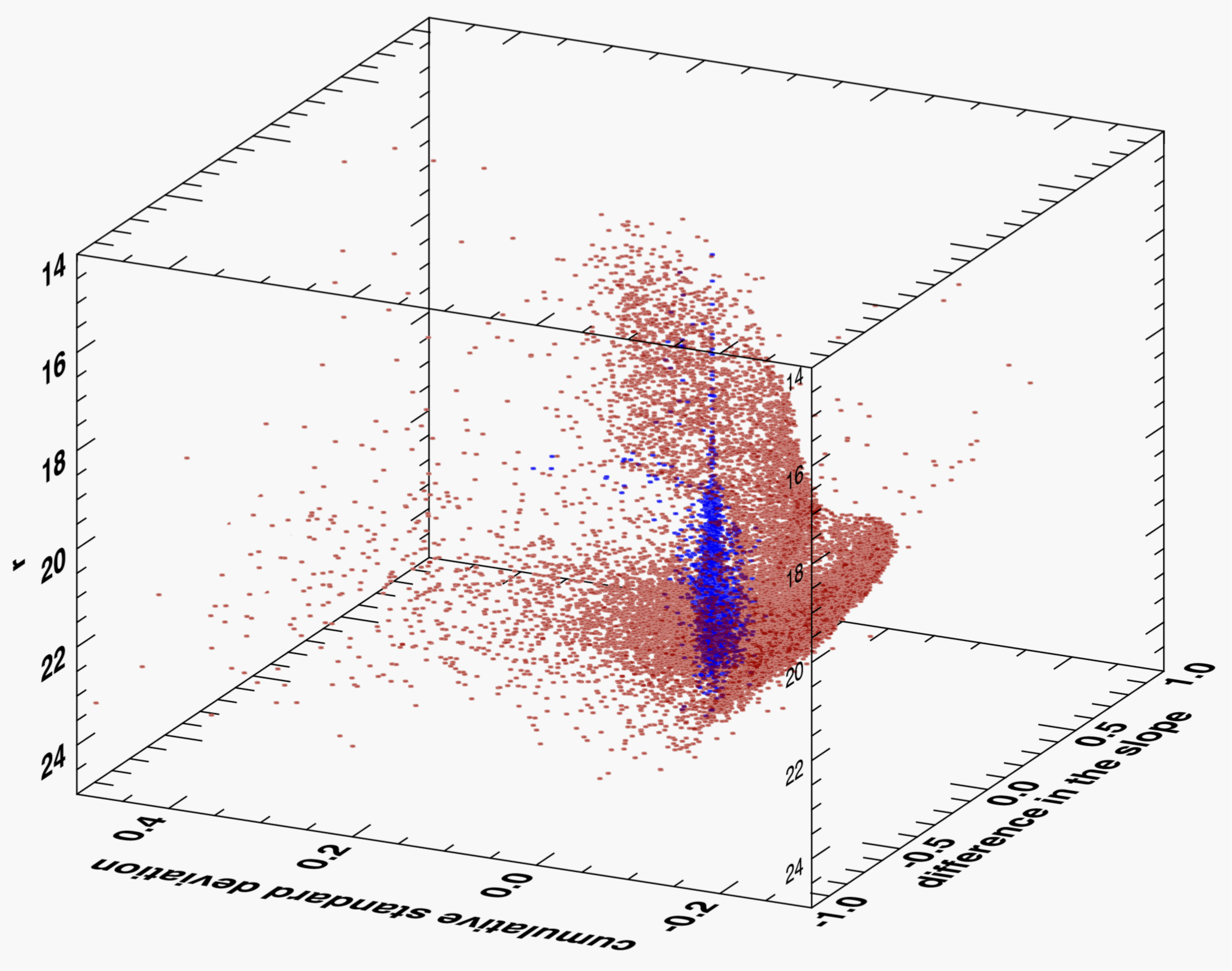}
\caption{3D plot showing the $r$-band magnitude as a function of the cumulative standard deviation
and of the difference in the slope. The color coding is the same as in
Fig. \ref{rgrgi}, blue
dots are candidate cluster stars, while pink dots are candidate field stars.}\label{select}
\end{figure}
The key steps are the following: we estimated the ridge lines in the
$r$,$g-r$ and $r$,$r-i$ CMDs tracing the cluster red-giant branch
(RGB), the main sequence (MS)  and the main sequence turn-off (MSTO).
The candidate horizontal branch (HB) stars were not included since
they are a small fraction of the cluster stellar population. The ridge
lines were estimated neglecting the stars located in the innermost
cluster regions ($r\le$ 1') and applying severe cuts in radial
distance and photometric accuracy. Fig. \ref{rgrgi} shows 
the resulting 3D ridge line (red line) and the initial
separation between candidate field (pink dots) and cluster (blue dots)
stars.

Subsequently, we performed a linear interpolation among the $r$,$g-r$
and $r$,$r-i$ ridge lines and adopted two different statistical
parameters to separate field and cluster stars:

a) we assigned a figure of merit to the difference in the slopes of
the three apparent magnitudes of each individual star and the three
magnitudes of the cluster ridge line;

b) we estimated the cumulative standard deviation among the magnitudes of
each individual star and the magnitudes of the cluster ridge line.

Then we generated a new 3D plot (see Fig. \ref{select}) in which
candidate field (pink dots) and cluster (blue dots) stars were plotted
as a function of the $r$-band magnitude, the difference in the slope and
the cumulative standard deviation. We estimated two new ridge
lines and the distance of individual objects from  them was eventually
adopted to provide the final separation between candidate field and
cluster stars.  Note that this approach was conservative, i.e.,
it was preferable to lose some possible candidate cluster members
rather than including any possible candidate field stars.

\begin{figure}
\includegraphics[width=\columnwidth]{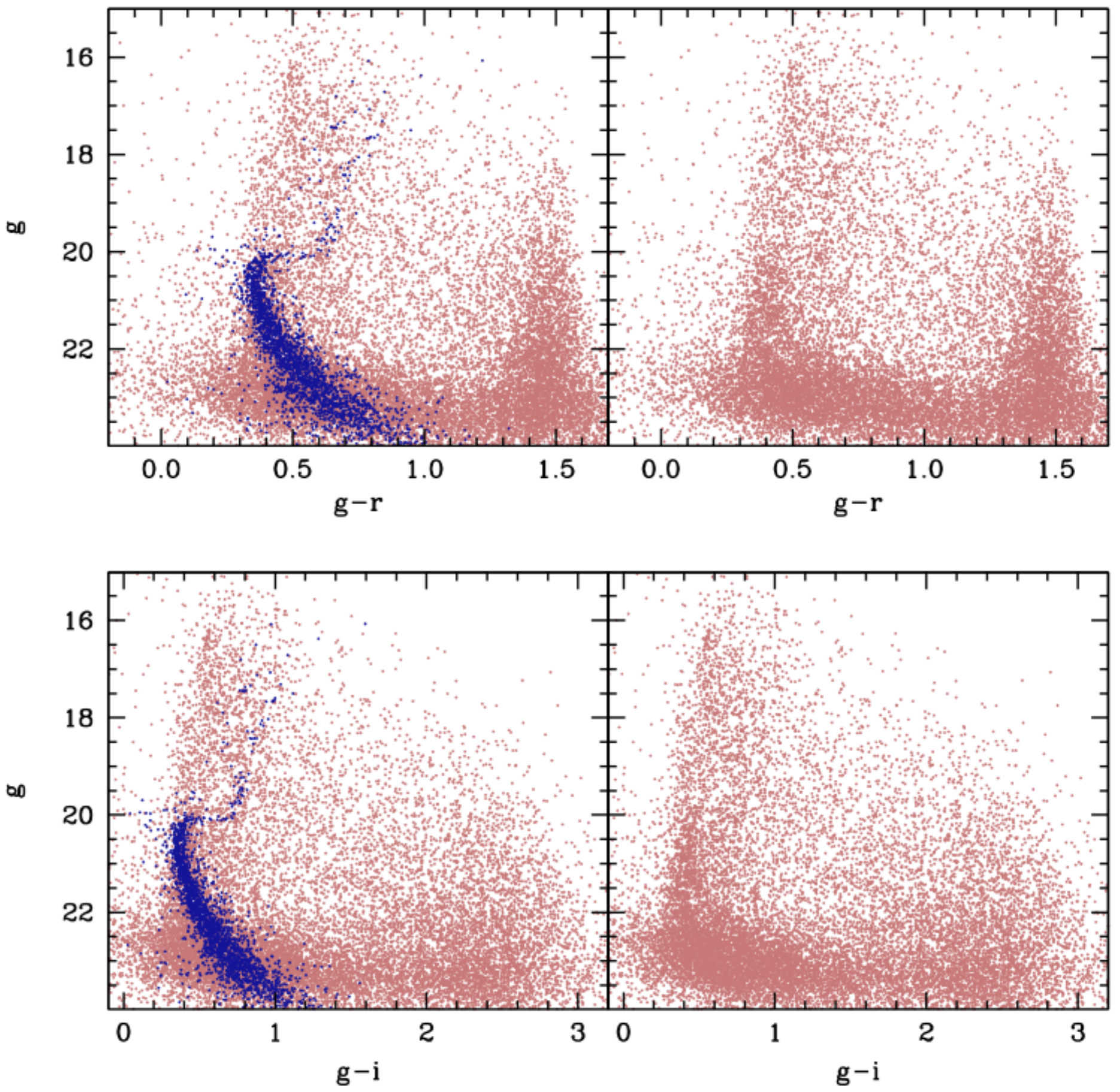}
\caption{The $g$,$g-i$ and $g$,$g-r$ CMDs are shown in the lower and
  upper panels, respectively. Pink and blue dots
  represent the field and cluster stars, respectively, as obtained by the ``3D
  procedure'' (see text and Figs. \ref{rgrgi} and \ref{select} for
details).}\label{confcmd_bono}
\end{figure}

Figure~\ref{confcmd_bono} shows the $g$,$g-i$ (lower panels) and
$g$,$g-r$ (upper panels) CMDs obtained with the 3D approach in which
candidate field and cluster stars were plotted as pink and blue dots,
respectively. The outcome for both filter combinations is the same
not only concerning the candidate cluster stars, but also the
magnitude and color distribution of field stars.

Fig.~\ref{confcmd_bono} points out the advantages in using this
multi-dimensional approach when compared with the classical approach
based on bi-dimensional ridge lines: i) data plotted in the left
panels show that the candidate cluster stars were properly selected
not only along the RGB and the MSTO, but also in the fainter portion
of the MS ($g\le$22 mag), i.e., in a region of the CMD in which field
and cluster stars overlap in classical optical CMDs; ii) the field
star magnitude and color distribution in the right panel is quite
homogeneous and does not show the typical gaps left around 
bi-dimensional ridge lines.

\begin{figure}
\includegraphics[width=0.95\columnwidth]{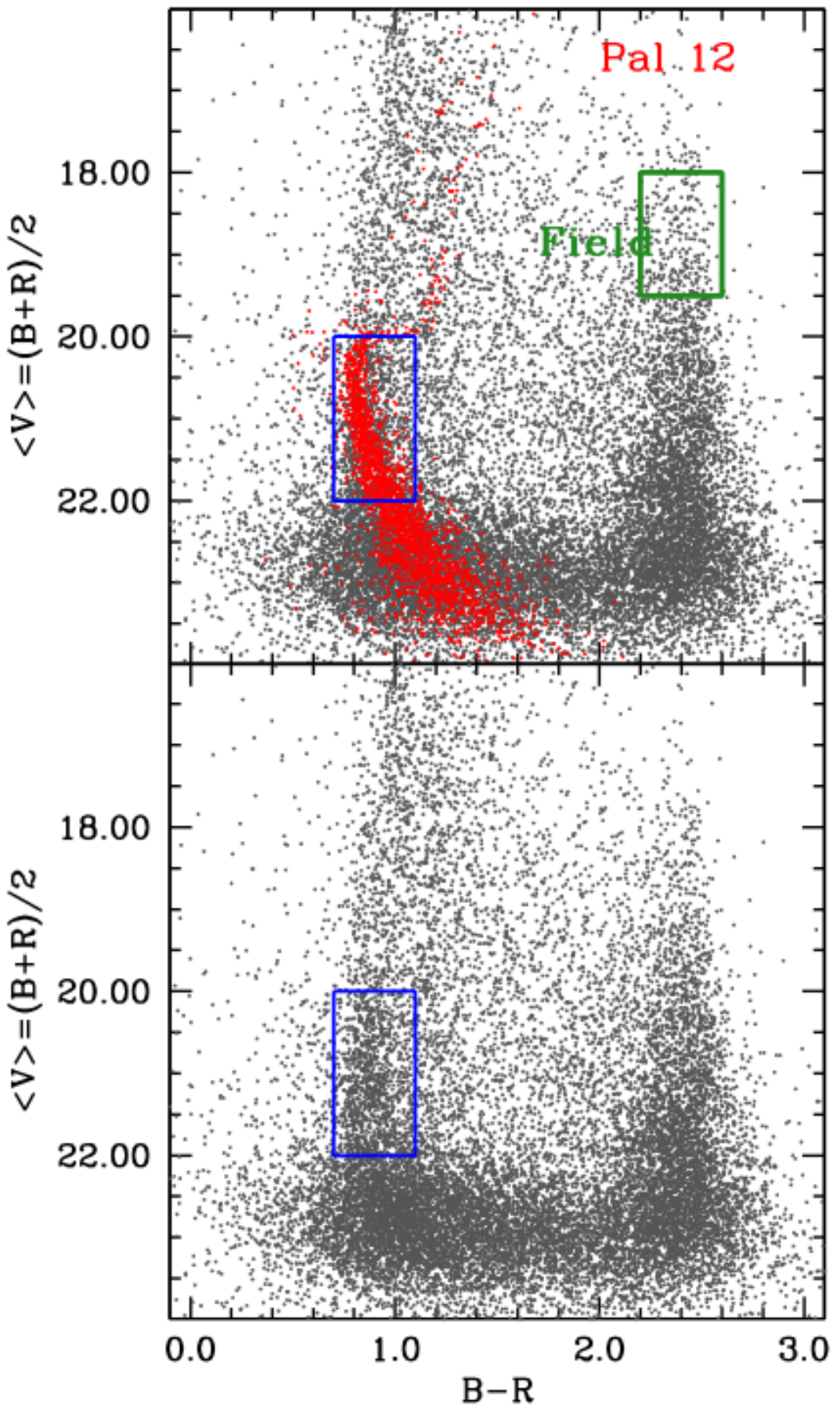}
\caption{$<V>$,$B-R$ CMD. In both panels, grey dots represent the
field stars obtained by the ``3D procedure''. In the upper panel the
red dots are the selected cluster member stars and the green
rectangle, labelled as {\it Field}, is a CMD region where we do not
expect stars belonging to Pal~12 or the Sgr stream. The blue rectangle in
the lower panel marks the overdensity CMD region. See text for
details.}\label{cmd_bono}
\end{figure}

In Fig. \ref{cmd_bono}, we report the $<V>,B-R$ CMD resulting from
the ``3D procedure''. In both panels, we show in grey the stars
recognised as field, and in the upper panel, we plot in red the
selected Pal~12 members. From inspection of the lower panel, excluding
the selected  Pal~12 members, the
structure with magnitude fainter than 20 mag and color ranging between
0.7 and 1.1 mag (blue rectangle) already noted in
Fig. \ref{cmd_MD} is clearly present. The green rectangle, labelled ``Field'' in the
upper panel of Fig \ref{cmd_bono}, selected as in Fig. \ref{cmd},
represents a CMD region where we do not expect stars belonging to
Pal~12 or the Sgr stream. It will be used for comparison in the following
analysis. 

\begin{figure}
\includegraphics[width=0.95\columnwidth]{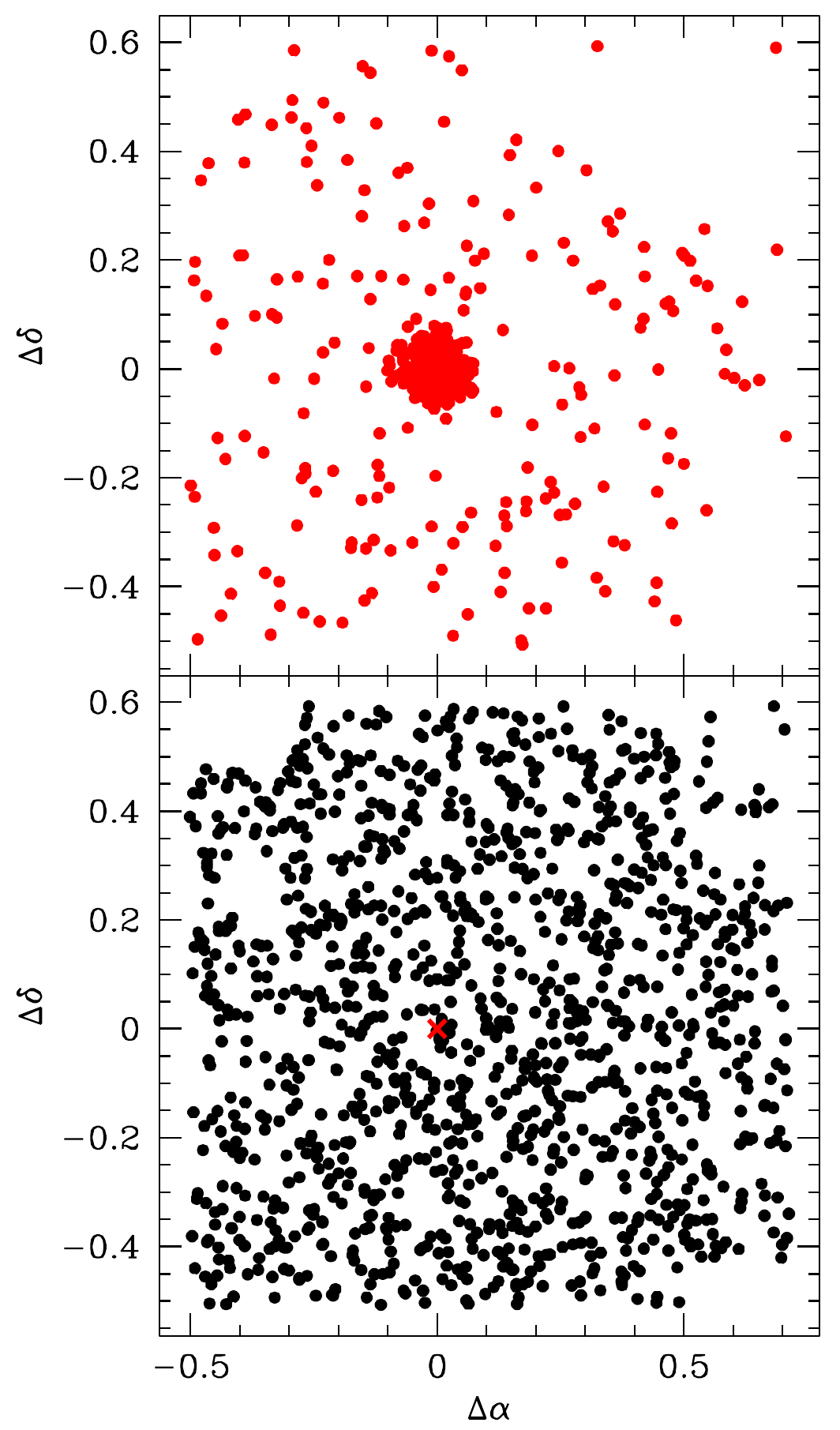}
\caption{The spatial distribution of the cluster (red dots in the
upper panel) and field stars (black dots in the lower panel) included
in the blue rectangle of Fig. \ref{cmd_bono}. The red cross in the lower
panel marks the Pal~12 center.}\label{xyoverd}
\end{figure}

The spatial distributions of the cluster (red dots) and field stars (black
dots) in the blue rectangle are shown in the upper and lower
panels of  Fig. \ref{xyoverd}, respectively (the red cross in the
lower panel marks the Pal~12 center). In the lower panel we do not
note any particular spatial overdensity and the comparison with the
cluster stars in the upper panel suggests that at most few Pal~12
members have been missed by the adopted ``3D procedure''.  For
this reason, in this structure we expect to have MW halo stars and,
taking also into account the results of \citet{MD02}, the overdensity
due to the Sgr stream.  Moreover, from this analysis it is clear that
the ``3D procedure'' appears to work very well to select Pal~12
members. On the other hand, with this approach it is very difficult
to distinguish Pal~12 from Sgr stream stars due to the  very similar
CMD characteristics. For this reason, we expect that a fraction of the
stream stars  are assigned to Pal~12 and a fraction to the field.

\begin{figure}
\includegraphics[width=0.95\columnwidth]{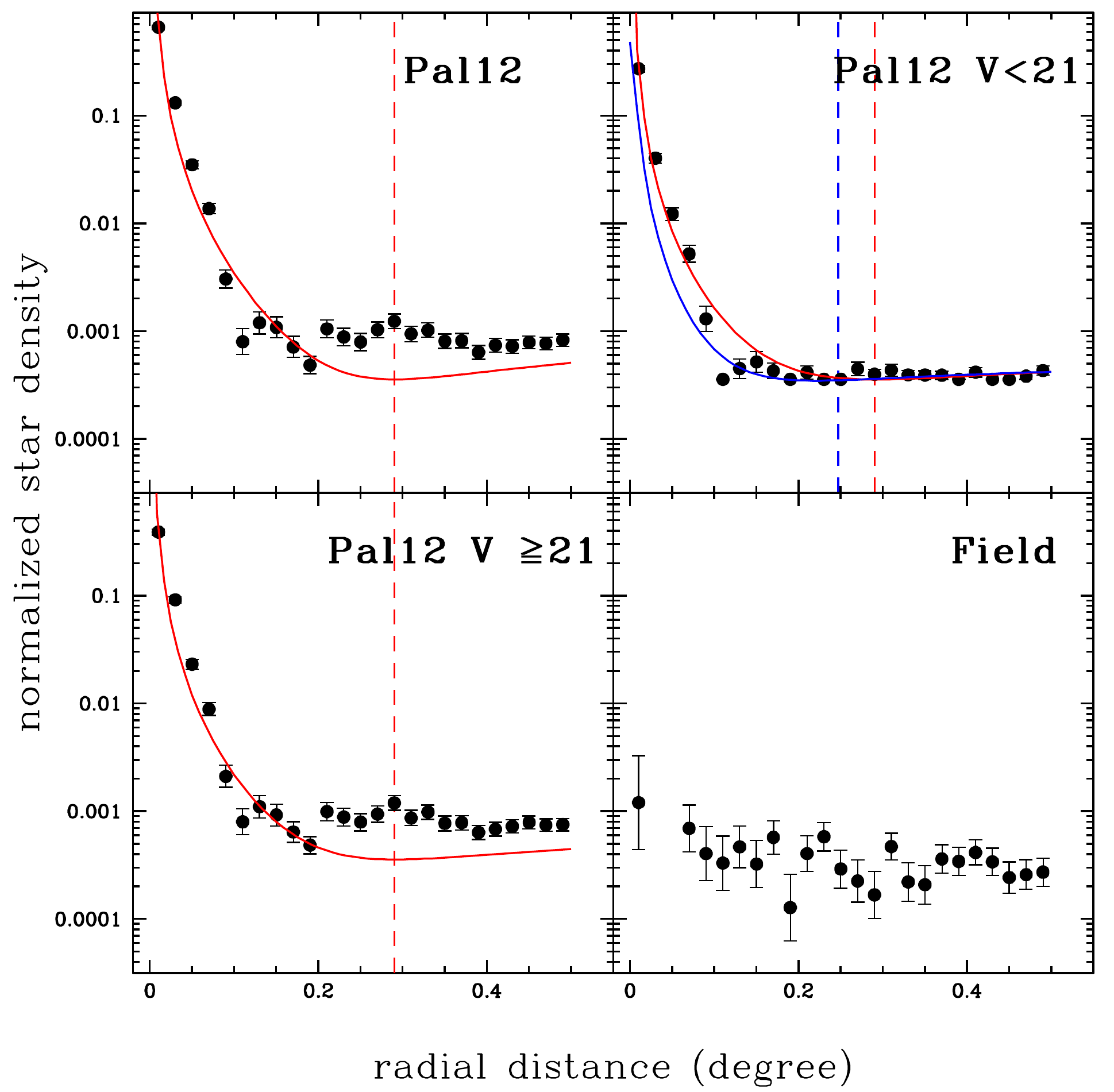}
\caption{Radial counts of the normalised star densities for  different
star selections in Fig. \ref{cmd_bono} (see text for details). The red
solid line represents the King profile of Pal~12 for core radius and
tidal radius by \citet{Harris96} and the red  vertical dashed line
marks this tidal radius.  The blue solid line
represents the King profile obtained in this paper from fitting the radial
counts for Pal 12 stars with $V<21$ mag, and the blue vertical dashed
line marks the derived tidal radius.}\label{cont_bono}
\end{figure}

In Fig. \ref{cont_bono}, we plot the radial counts of the
normalised star densities for the different star selections in
Fig. \ref{cmd_bono}: i) in the left-upper panel, we show the red
stars, labeled ``Pal~12'', ii) in the right-upper and in the
left-bottom panels, are the results obtained using only the stars with
$V<21$ and $V \ge 21$ mag, respectively and iii) in the right-bottom panel
are the star counts for the ``Field'' stars.  To obtain these counts, we
adopted the procedure described above. A comparison between
Figs. \ref{cont}  and \ref{cont_bono} shows that the selection of the
cluster stars by the 3D procedure produces smaller Poisson errors
(also in this case, we added to the radial counts the average of the
``Field'' counts).  For all the star selections in
Fig. \ref{cont_bono} we obtain a profile typical of a GC
 \citep[see e.g.][]{King66,Wilson75}. ``Field'' stars, as
expected, are the same as in Fig. \ref{cont}. Also in this case, the
red line represents the King profile for the nominal value of Pal~12's
core radius and tidal radius. For the two selections in the left
panels (including the stellar overdensity), the normalised star density
shows the expected excess of stars around the nominal tidal radius
(vertical red dashed line) due to the stars belonging to the Sgr
stream that our procedure cannot distinguish from Pal~12 members.
For  Pal~12 stars with $V<21$ mag, we have an almost negligible
overdensity around 0.29$\,$deg, and the nominal tidal radius seems to be
too high. Fitting a King profile to these data we obtain the blue
solid line with $c=1.68$, $r_c=0.28$ arcmin and $r_T=13.2$ arcmin
$=0.22$ deg (vertical blue dashed line). On the basis of these
results, the presence of the Sgr stream seems to have mimicked a
larger tidal radius. However we note that for $V<21$ the number of
stars is too small to get firm constraints on the King profile
parameters.

\section{Conclusions}\label{sec-conclusions}

In this paper, we have studied the LF and the star counts in a region
around the globular cluster Pal~12, covering about two tidal radii.
We have compared our analysis with previous studies that identified a
CMD star overabundance, a possible signature of the Sgr stream from
which Pal~12 was probably originated.  Our study confirms the
presence of a stellar overdensity at $<V>$ fainter than $\sim 20.5$\,mag
and $B-R$ color ranging between 0.6 and 1.0 mag. The analysis of the overdensity's
LF and CMD suggests a dominant stellar population older than 10 Gyr
and with a mean metallicity $[Fe/H]\sim -1$ dex, consistent with the
old stellar components in the Sgr dwarf galaxy.  Most of these
stars appear to be located  at the distance of Pal 12 ($\sim 18.2$
kpc), with a possible fraction of stars at shorter/longer distances
(down to $\sim$17 kpc/up to the Sgr distance).

The presence of the Sgr stream affects our evaluation of star counts
as a function of the radial distance from the center of Pal~12.  This is
more evident when  adopting  the ``classical'' (bi-dimensional)
approach of selecting the stars in a magnitude range around the ridge
line in a specific CMD, to identify Pal~12 candidate members. On the
other hand, the use of an innovative procedure based on a 3D ridge
line (magnitude-color-color) allowed us to perform a better separation
between Pal~12 and field stars and to significantly reduce the
contamination from the Sgr stream.

We do not find evidence of significant extra-tidal Pal~12 stellar
population. On the contrary, the presence of the Sgr stream
might have simulated a larger tidal radius in previous studies.

\section{Acknowledgments}

This work has made use of BaSTI web tools. Partial financial support
for this work was provided by PRIN-INAF 2011 Tracing the formation
and evolution of the Galactic halo with VST (PI: M. Marconi),
PRIN-INAF 2011 Galaxy Evolution with the VLT Surveys Telescope (VST)
(PI: A. Grado)  and  PRIN-INAF 2014 project EXCALIBURS (PI: G. Clementini).

{}

\end{document}